\documentclass{aa520}
\usepackage{natbib}
\usepackage{times,graphics,latexsym,amssymb}
\bibpunct{(}{)}{;}{a}{}{,}
\begin{document}

\title{Stochastic chemical enrichment in metal-poor systems}
\subtitle{II. Abundance ratios and scatter}
\author{T. Karlsson and B. Gustafsson}
\offprints{T. Karlsson (Torgny.Karlsson@astro.uu.se)}
\institute{Department of Astronomy and Space Physics, Box 515, SE-751 20, 
Uppsala, Sweden}
\titlerunning{Stochastic chemical enrichment}
\abstract{
\noindent
A stochastic model of the chemical enrichment of metal-poor systems by core 
collapse supernovae is used to study the scatter in stellar
abundance ratios. Large-scale mixing of 
the enriched material by turbulent motions and cloud collisions in the 
interstellar medium, and infall of pristine matter are taken into account. 
The resulting scatter in abundance ratios, e.g. as functions of the overall
metallicity, is demonstrated to be crucially dependent on the as yet uncertain
supernovae yields. The observed abundance ratios and their scatters therefore 
have diagnostic power as regards the yields. The relatively small star-to-star 
scatter observed in many chemical abundance ratios, e.g. by Cayrel et al. 
(2004\nocite{cayrel04}) for stars down to $[\mathrm{Fe}/\mathrm{H}]= -4$,
is tentatively explained by the averaging of a large number of contributing 
supernovae {\it and} by the cosmic selection effects favoring contributions from
supernovae in a certain mass range for the most metal-poor stars. The scatter
in observed abundances of $\alpha$-elements is understood in terms of 
observational errors only, while additional spread in yields or sites of
nucleosynthesis may affect the odd-even elements Na and Al. For the iron-group
elements we find some systematic deviations from observations in abundance 
ratios, such as systematically too high predicted Cr/Fe and Cr/Mg ratios, as well
as differences between the different sets of yields, both in terms of 
predicted abundance ratios and scatter. The semi-empirical yields recently
suggested by Francois et al. (2004\nocite{francois04}) are found to lead
to scatter in abundance ratios significantly greater than observed, when
applied in the inhomogeneous models. 
"Spurs", very narrow sequences in abundance-ratio diagrams, may disclose 
a single-supernova origin of the elements of the stars
on the sequence. Verification of the existence of such features, called single 
supernova sequences (SSSs), is challenging. This will require samples of several
hundred stars with abundance ratios observed to accuracies of 0.05 dex or 
better.
}
\maketitle

\section{Introduction} \label{intro}
\noindent
During the recent decades progress in stellar spectroscopy and abundance 
analysis has made it possible not only to explore {\it trends} in
abundance ratios with varying overall metallicity, for various elements
and stellar populations, but also to study the possible presence and
origin of significant cosmic {\it scatter} in these ratios. In many cases this 
scatter is smaller than $0.2$ dex in the logarithmic ratios and requires
detailed and accurate, preferably strictly differential, analysis in order to 
be studied. One example of such a study is the survey by Edvardsson et al. (1993\nocite{edvardsson93})
of solar-type stars in the Galactic disk; for these the scatter in 
alpha-element (Mg, Si, Ca) abundances relative to iron was found to be very small
($0.05$ dex or less) at a given iron abundance [Fe/H],
while the scatter in [Fe$/$H] at a given stellar age and
characteristic distance from the Galactic center was on the order of $0.2$ dex.
For a sample of Pop II stars Nissen et al. (1994\nocite{netal94}) concluded from the small 
scatter ($<\!0.06$ dex) in alpha-element abundances relative to iron, 
that the number of supernovae (SNe) that had previously enriched the gas out of which
the observed stars were once formed should have been on the order of $20$ or more. 
These conclusions were based on the fact that theoretical yields of different
heavy elements from models of core-collapse SNe depend strongly, among 
other things, on the progenitor mass of the SN 
(e.g. Woosley \& Weaver 1995\nocite{ww95}). Thus, a limited number of 
SNe, drawn randomly from the initial mass function (IMF), should inevitably lead
to a scatter in relative abundances for stars in the following generation
of different origin but similar
low overall metallicity. This was also pointed out by Audouze and
Silk (1995\nocite{as95}) who estimated that due to this stochastic effect, 
significant scatter should be found among stars with $[\mathrm{Fe}/\mathrm{H}]<-2.5$.

\par

A first model of stochastic nucleosynthesis in the Galactic halo
was developed by Tsujimoto et al. (1999\nocite{tsy99}).
Argast et al. (2000\nocite{argast00}) and Oey (2000\nocite{oey00}) also developed stochastic Halo 
formation models. These models are based on simple assumptions about the mixing of 
enriched SN gas with the surrounding interstellar medium (ISM) as regards masses of gas involved 
and time scales for the process. They differ in important details --
e.g. Argast et al. (2000\nocite{argast00}) considered individual SNe as primary elements in their 
model while the model of Oey (2000\nocite{oey00}) is based on multi-SN superbubbles. 
Technically, the 
models are also different -- e.g. the model of Argast et al. (2000\nocite{argast00}) is basically 
numerical while that of Oey (2000\nocite{oey00}) is partially developed analytically. In none
of these cases are the models dynamically self-consistent -- the mixing of gas with the
surroundings is described by recipes that may be reasonable, with some
support in other theoretical studies or observations, but they are based on
ad-hoc assumptions and contain several free parameters. 
These models have been used in different applications with comparisons to 
observations. Tsujimoto et al. (1999)\nocite{tsy99} applied their
model to interpret the observed distribution of  Galactic halo
stars in the [Eu$/$Fe]-[Fe$/$H] diagram with a scatter tending to increase towards 
the low metallicities. The yields for the r-process
element Eu were, and had to be, more or less assumed; i.e. they were not
based on detailed nucleosynthesis calculations for SNe. Oey (2003\nocite{oey03}) compared 
results from her model to the observed metallicity distribution function of 
Galactic halo stars and applied it to 
predict the number fraction of zero-metallicity (Population III) stars. Argast
et al. (2000\nocite{argast00}) calculated the chemical inhomogeneous enrichment of the ISM
caused by single core-collapse supernovae and studied the resulting 
element abundance patterns and scatter in abundance ratios expected for stars
of different metallicities. For [Fe$/$H] $< -3.0$ they found the ISM
to be still essentially
unmixed and dominated by local inhomogeneities. For 
$-3.0\leq$ [Fe$/$H] $\leq -2.0$ the
mixing increased gradually, and when the gas became more metal-rich it became well
mixed. These authors were able to reproduce the observed scatter among Pop II
stars for some elements, Si, Ca and Eu, relative to Fe, while their calculated
scatter in O and Mg relative to Fe is too large, and in Ni too small. They
explained these discrepancies in terms of errors in the stellar yields adopted.

\par

The relation between the yields adopted and
the resulting abundance scatter was further explored and modelled in 
a statistical and partially analytical model of inhomogeneous enrichment by 
Karlsson \& Gustafsson (2001\nocite{kg01}, hereafter KG). We demonstrated 
how the yields are mapped into the relative-abundance diagrams by simple 
multiple integrals, where the integration, however, has to be carried out 
over complicated regions in mass-space determined by the yields. In a more recent paper, 
Karlsson (2005a\nocite{paperi}, hereafter Paper~I) 
has developed this model further, derived an explicit form of the time-development
of the mixing-volume surrounding each SN in the ISM and added the effect
of (space-independent) infall of pristine gas. This model departs from those
of Oey (2000\nocite{oey00}) and Argast et al. (2000\nocite{argast00}) in a number of details, which will
be further commented on below, but it also has a number of features in common 
with those models. We shall here use the model to further study the 
relative-abundance diagrams expected for the most metal-poor stars. Argast
et al. (2000\nocite{argast00})  concentrated on the abundances of different elements 
relative to iron. The iron yields, however, are uncertain since they are
directly dependent on the uncertainties in the explosion mechanisms and the
size of the collapsing iron core. We shall instead mainly discuss trends and 
scatter of stars in diagrams such as [Si$/$Mg] vs. [Mg$/$H] or [Ca$/$Mg]. Although the 
yields of these elements may be considered to be more certain, we shall use different
published yields from the literature to explore the differences in the 
predicted distributions in the relative-abundance diagrams. We find that
the resulting distributions are highly sensitive to the adopted yields as 
functions of stellar mass. When comparing to recent observational results for
Extreme Pop II stars, we will see that important restrictions are put on 
the yields by the observations. 
Recent observational studies indicate that the star-to-star 
scatter is surprisingly small in many abundance ratios, excluding the 
neutron-capture elements, even for stars with a metallicity as low as 
$[\mathrm{Fe}/\mathrm{H}]=-4$ (Carretta et al. 2002\nocite{carretta02}; 
Cayrel et al. 2004\nocite{cayrel04}).
One result of the present study will be that the
astonishingly small dispersion in many abundance ratios among the most
metal-poor stars has a natural explanation within the framework of the 
stochastical model -- e.g., extra global mixing in the early ISM does not have
to be invoked.

\par

In Sect. \ref{model}  of this paper we sketch the present nucleosynthesis model 
and the yields adopted. In Sect. \ref{results} a number of results are 
demonstrated and compared with observations. In Sect. \ref{conclusions} 
conclusions are made and possible implications 
for future observations and theoretical efforts are discussed.  

\section{Models of the early chemical evolution}\label{model}
\noindent
Our method of modelling the chemical evolution of the early Galaxy, and the resulting chemical abundances in Pop II stars is described in detail in Paper~I\nocite{paperi}. Here, only a rough sketch will be given for reference.

\par

In order to model the built-up of chemical elements in the early ISM and link it to the abundances observed in low-mass stars we require knowledge of

\begin{itemize}
\item[(1)] when, where, and at what rate SNe and other sites of 
           nucleosynthesis are active,
\item[(2)] what elements and how much of each are ejected from each site,
\item[(3)] how the ejected material is dispersed,
\item[(4)] to which extent, where and when infall of extra-galactic gas 
           affects the composition of the ISM, and
\item[(5)] when and where the local abundance is probed by subsequent star 
           formation.
\end{itemize}

\par

We shall here shortly comment on the underlying assumptions and the main
features in our modelling of the various aspects listed. Since the 
SN rate and the star formation are closely linked, we shall first discuss
(1) and (5) above.
 
\subsection{Star formation and sites of nucleosynthesis}
\noindent
Star formation is a clustered phenomenon and many stars are formed in groups 
like globular clusters, open clusters and OB associations. However, there is 
also less clustered star formation and most of the field stars were presumably not
formed in strongly gravitationally bound associations (see, e.g., 
Oey et al. 2004\nocite{okp04}). A fundamental assumption in the present model 
is that star formation is unclustered and that the stars are randomly distributed 
in space (cf. Karlsson 2005b\nocite{paperiii}, hereafter Paper~III). 
This means that the star formation rate (SFR) can be described by a function 
$\psi=\psi(t)$, which depends only on time. Under this assumption, chemical 
enrichment events, such as SNe, are spatially uncorrelated, as is the mapping 
of the chemical inhomogeneities by the following generations of low-mass stars. 

\par

The SFR in Paper~I\nocite{paperi} was alternatively assumed to be 
constant with time during the evolution of the Galactic halo, or based on the numerical 
simulations of Samland et al. (1997\nocite{setal97}).
Their SFR for the Halo shows a sharp rise at early times, a maximum around 
$2$ Gyr followed by a 
rather steep decline which gradually levels off at around $3$ Gyr. In Paper I  
some consequences of various assumptions concerning the SFR were explored.
Having found that our resulting relative abundances are only very weakly
dependent on the assumptions we make for the SFR (abundances relative 
to hydrogen are, however, less insensitive to the SFR, see Sect. \ref{pardep}) 
we have here adopted it as constant during $1$ Gyr, with a value of $125$ 
stars kpc$^{-3}$ Myr$^{-1}$ (i.e. Model A, in Paper~I).

\par

As we are aiming at modelling only the early phases in the Galaxy, where
core-collapse supernovae are thought to be the only objects with time-scales 
short enough to contribute importantly to nucleosynthesis 
(see Paper~I\nocite{paperi}), we have only taken those sites into 
consideration here. We have assumed the SN rate to scale with the 
star-formation rate. That is, the time delay between star formation and 
the end of massive-star evolution has been neglected. We have assumed
all stars in the mass interval $8-100~\mathcal{M_{\odot}}$ 
to explode as core-collapse SNe. No extrapolations of yields were, 
however, made beyond the mass limits of the tables by different authors. 
The initial masses of the individual SNe are drawn randomly, distributed 
according to the Salpeter IMF.

\subsection{Yields}
\noindent
A strong dependence of calculated SN yields on initial stellar 
mass, different for different chemical elements, implies a scatter 
in relative abundances for gas enriched by a small number of supernovae.  
However, the uncertainties in the supernova models lead to considerable
differences between the results of the different sets of yield calculations. 
A summary of these uncertainties is given by Argast et al. (2000\nocite{argast00}). 
Three different sets of core-collapse yields, calculated in a metal-free 
environment, have been used here -- those of Woosley \& Weaver (1995\nocite{ww95}, 
hereafter WW95), those of Umeda \& Nomoto (2002\nocite{un02}, hereafter UN02) 
and those of Chieffi \& Limongi (2004\nocite{cl04}, hereafter CL04). 
For the yields by WW95 we have used the models Z12A, Z13A, Z15A, Z22A,
Z25B, Z30B, Z35C, and Z40C, compensating for the radioactive decay of 
unstable nuclei. For the yields by UN02 we have used the models with 
explosion energy $E=1\times 10^{51}$ erg. 

\par

These different sets of calculations differ in important
nuclear cross sections, the treatment of convection and mixing, and the
choice of explosion energy and mass of the collapsing iron core. It is 
a complex task to analyse how the yields are affected by the different
choices (CL04) and will not be discussed here.

\subsection{Dispersion of supernova products, mixing in ISM}
\noindent
A key concept in our model of early enrichment of the ISM is 
the mixing volume $V_{\mathrm{mix}}=V_{\mathrm{mix}}(t)$. One mixing volume is 
assumed to be generated by each SN and is defined as 
the total volume enriched by debris from this SN,  
including the region directly
affected by the explosion as well as those regions which are reached
due to the turbulent motions in the ISM. As time goes on, the mixing volumes 
begin to overlap in space. Chemical inhomogeneities are developed 
by subsequent generations of SNe that explode within earlier generations of 
mixing volumes and enrich the gas further.  At any instant in time, different 
regions of the ISM will be enriched by 
different numbers of SNe and there exists a distribution 
$w_{\mathrm{ISM}}(k,t)$ describing the probability of finding $k$ overlapping 
mixing volumes at time $t$ somewhere in the ISM. Via this distribution, we 
are able to quantify and map the abundance distributions in stars. These abundance 
distributions, on the other hand, may be regarded as transformations of
the distributions of supernova masses, via the supernova yields. The calculation
of these transformations is elaborated in KG and summarized in 
Appendix \ref{app_derivation}, below.

\par

The form of $w_{\mathrm{ISM}}$ may be inferred from the spatial Poisson process. Let $N$ be a number of points, randomly distributed in a box of size $V$. The probability of finding exactly $k$ points in a small volume $V'$ ($V'\ll V$) is then approximated by a Poisson distribution $P(k,\mu')$, where the parameter $\mu'$ is defined as $\mu'=V'\times (N/V)$ (cf., e.g., Cram\'{e}r 1945). On the other hand, if the points in the box are replaced by volumes of size $V''$, the probability of finding $k$ overlapping volumes is again given by a Poisson distribution $P(k,\mu'')$, with $\mu''=(V''\times N)/V$. Evidently, the two probability distributions are identical for $V''=V'$. The latter situation is equivalent to the the picture described above, where the ejecta from SNe are spread over finite volumes $V''=V_{\mathrm{mix}}$. Hence, assuming that the SNe are randomly distributed over space with an average rate $u_{\mathrm{SN}}$ per unit time and unit volume, we may set the probability of finding a region enriched by $k$ SNe to be

\begin{equation}
w_{\mathrm{ISM}}(k,t)=P(k,\mu(t))=e^{-\mu(t)}\mu(t)^{k}/k!,
\label{poisson}
\end{equation}

\noindent
where $\mu(t)$ denotes the integrated volume affected by SNe expressed in units of the total volume of the system, at time $t$.

\par

Dispersive processes like turbulent diffusion will gradually spread the SN material over larger and larger volumes. More precisely, if, at time $t$, $V_{\mathrm{mix}}(t-t')$ is the size of the volumes enriched by material from SNe that exploded at time $t'$ then

\begin{equation}
\mu(t) = \int\limits_{0}^{t} V_{\mathrm{mix}}(t-t')u_{\mathrm{SN}}(t')\mathrm{d}t'.
\label{muoft}
\end{equation}

\noindent 
With this definition, $\mu(t)$ is a measure of the average number of SNe that 
have enriched a random volume element in space at time $t$. The chemical 
inhomogeneities should be well described by Poisson statistics as long as the 
SNe are assumed to be randomly distributed in the ISM and $\mu\ll N_{\mathrm{SN}}$, the total number of SNe in the system.

\par

The different evolutionary phases of a SN remnant are well understood. However, this initial spread of the SN material is confined within a small volume and the chance of immediately forming stars in the swept-up gas is probably not very great. The successive, large-scale mixing of the cooled SN material, caused for example by turbulent motions, must be included. Following Bateman \& Larson (1993)\nocite{bl93} and neglecting the relatively small initial expansion of the SN remnant itself, we estimate that the mixing volume should on average increase as

\begin{equation}
V_{\mathrm{mix}}(t)=\frac{4\pi}{3}R_{\mathrm{rms}}^3=\frac{4\pi}{3}\left(\sigma
_{\mathrm{mix}}t \right)^{3/2}.
\label{vmix}
\end{equation}

\noindent
Here $t$ denotes the time after the SN explosion and 
$\sigma_{\mathrm{mix}}$ is a combined cross-sectional expansion 
rate for the processes responsible for the bulk 
distribution of heavy elements. From the data given in 
Bateman \& Larson (1993\nocite{bl93}) we 
find \mbox{$\sigma_{\mathrm{mix}}=7\times 10^{-4}$ kpc$^2$ Myr$^{-1}$}. 

\par

From Eqs. (\ref{poisson}) -- (\ref{vmix}) we may now 
derive analytical expressions for the distribution $f_{M_k}$ of mixing 
masses and the number of persisting stars formed by gas polluted by $k$ 
different supernovae, $n_k$, with $k$ ranging from zero to 
high numbers (see Appendix \ref{app_nkfmk}). The mixing mass is the 
total amount of gas within the mixing volume $V_{\mathrm{mix}}$, in 
which the newly synthesized elements are mixed. For the mixing masses, 
we find rather narrow distributions extending from about $10^5$ to 
$10^6~\mathcal{M_{\odot}}$ for $k=1$. The widths of the distributions 
for $k=100$ are on the order of $5-10$ times larger, depending on the 
input parameters (see Appendix \ref{app_nkfmk}, Fig. \ref{Mmix_fig} or Paper I\nocite{paperi}, Fig. 3). 
Taking the high-mass IMF with the yields for each SN initial mass, we may 
now also for each $k$ calculate the expected 
distribution of stellar abundances of any chemical element, $A$.
A summation over $k$ leads to the final distributions of stars with different
abundances and abundance ratios. This, however, also requires a consideration
of the infall of extra-galactic gas during the formation of the Halo.

\subsection{Infall of extra-galactic gas}
\noindent
The gas density in our model plays an obvious role in the calculation of
the abundances resulting from adding supernova processed matter into the
ISM. Just as for the star-formation rate $\psi=\psi(t)$, we assume a 
space-averaged density $\rho(t)$ to be representative and may easily write a
differential equation for the time-derivative of $\rho(t)$ 
which includes sinks due to star formation
and sources due to ejection of gas (remaining in the system) and accretion
of gas from extra-galactic space. We assume the latter to be pristine, i.e.
composed only of hydrogen and helium. In this study we assume a constant gas density 
(i.e., Model A, discussed in Paper~I). This corresponds to the extreme infall
model by Larson (1972), where the infall compensates for the mass locked up in stars and
stellar remnants. With the adopted SFR, the average infall rate is 
$60~\mathcal{M_{\odot}}$~kpc$^{-3}$~Myr$^{-1}$ or approximately $16~\mathcal{M_{\odot}}$~yr$^{-1}$ 
for all the Galaxy, during the first billion years. In our space-averaged 
model for the gas density, this infall 
is assumed to be diffuse, i.e. not directly leading to localized chemical inhomogeneities.
We note, however, that a generalization of our model to allow for a 
stochastic inhomogeneous infall in distinct clouds should be possible. 

\begin{center}
\begin{table}[t]
\caption{Parameter dependence on $[A/\mathrm{H}]^{\mathrm{a}}$}
  \label{pardep_tab}
  \begin{tabular}{lcll}
     \hline
     \hline
     \\*[-0.5em]
\hfill\footnotesize{Parameter}\hfill{} & \hfill\footnotesize{Incr.}\hfill{} & \hfill\footnotesize{Equiv. to}\hfill{} & \hfill\footnotesize{$\langle [A/\mathrm{H}]_{\mathrm{new}}-[A/\mathrm{H}]_{\mathrm{old}}\rangle^b$}\hfill{} \\
     \\*[-0.5em]
     \hline
     \\*[-0.8em]
\footnotesize{$\overline{\rho}$} & \footnotesize{$\times 10$} & \footnotesize{$~~~~~~~~-$} & \footnotesize{$~~~~~~~~~~~~~-1$} \\*[0.1em]
 \footnotesize{$\overline{u}_{\mathrm{SN}}$} & \footnotesize{$\times 10$} & \footnotesize{$\overline{\rho}\times 10^{-3/5}$} & \footnotesize{$~~~~~~~~~~~~~+0.6$} \\*[0.1em]
 \footnotesize{$\sigma_{\mathrm{mix}}$} & \footnotesize{$\times 10$} & \footnotesize{$\overline{\rho}\times 10^{3/5}$} & \footnotesize{$~~~~~~~~~~~~~-0.6$} \\*[0.1em]
\footnotesize{$\overline{p}_A$} & \footnotesize{$\times 10$} & \footnotesize{$\overline{\rho}\times 10^{-1}$} & \footnotesize{$~~~~~~~~~~~~~+1$} \\*[0.1em] 
     \\*[-0.8em]
     \hline
     \\*[-0.8em]
  \end{tabular}

\hspace*{5pt}\footnotesize{$^{\mathrm{a}}$I.e., on $f_{M_k}$, for constant SFR and gas density of the ISM}  \\*[0.05em]
\hspace*{5pt}\footnotesize{$^{\mathrm{b}}$Average change} \\
\end{table}
\end{center}

\subsection{Model-parameter dependence on abundance ratios}
\label{pardep}
\noindent
We shall briefly comment on how abundance ratios depend on the different model parameters. We will distinguish between relative abundance ratios $[A/B]$ and absolute abundance ratios $[A/\mathrm{H}]$ measured relative to hydrogen, where $A$ and $B$ are two elements produced and ejected in SN explosions and $\mathrm{H}$ denotes hydrogen. Relative abundance ratios like $[A/B]$ are obviously strongly dependent on the yields of element $A$ and $B$. E.g., if the averaged yield $\overline{p}_A$ is increased a factor of ten, the abundance ratio is increased $1$ dex. However, although a single relative abundance ratio, e.g., measured in a point in the ISM, depends on the mixing conditions in the ISM, a collection (distribution) of relative abundance ratios (assembled from measurements of many points) is only very weakly dependent on the mixing (cf. KG). For abundance ratios measured relative to hydrogen the situation is different. Absolute abundance ratios like $[A/\mathrm{H}]$ depend, apart from the yield, on the amount of dilution of the SN ejecta, i.e., they depend on the mixing-mass distribution. The mixing-mass distribution $f_{M_k}$ depends, in turn, on the SFR and the density of the ISM, as well as the mixing velocity $\sigma_{\mathrm{mix}}$ via the expression for $\mu$ (see Appendix \ref{app_nkfmk}). In particular, for a model with constant SFR and gas density (e.g., Model A), the parameter dependences may be calculated exactly and are shown in Table \ref{pardep_tab}. The absolute abundance ratios depend linearly on the inverse of the gas density. An increase in the density of a factor of ten thus corresponds to a decrease in $[A/\mathrm{H}]$ of $1$ dex. On the other hand, a similar increase in the SFR results in an increase in $[A/\mathrm{H}]$ of $0.6$ dex. Due to the fact that the ISM is enriched by the same number of SNe on a shorter time-scale, the mixing volumes, and therefore the mixing masses, for any number $k$ of enriching SNe will be smaller, which results in a higher $[A/\mathrm{H}]$. A similar argument holds for the dependence on $\sigma_{\mathrm{mix}}$. Note that the change in $[A/\mathrm{H}]$ caused by a change in the SFR or $\sigma_{\mathrm{mix}}$ may be translated into a change in gas density, as shown in Table \ref{pardep_tab}. For constant models like Model A, different sets of input parameters will merely cause horizontal shifts of the density functions in diagrams relating a relative abundance ratio $[A/B]$ to an absolute abundance ratio $[A/\mathrm{H}]$.       
\par

For time-dependent SFRs and/or gas densities, the change in $[A/\mathrm{H}]$ ratios may not be calculated as easily as for models with constant $\psi$ and $\rho$. However, the general behavior is the same. Furthermore, an increasing gas density leads to a compression of the $[A/\mathrm{H}]$-scale, while a SFR which increases with time leads to a stretching of the $[A/\mathrm{H}]$-scale. Since the SFR is assumed to depend on the gas density of the medium ($\psi \propto \rho^{x}$, where $x$ is chosen between $1$ and $2$) the total effect of a varying gas density may thus be less than anticipated, due to the opposite behavior of the dependences on these two parameters. Note that $[A/B]$-scales are much less sensitive to compression and stretching caused by variations in the SFR/gas density than $[A/\mathrm{H}]$-scales. This is discussed in detail by KG\nocite{kg01}. Finally, the effect due to relative changes in the number $n_k$ of stars enriched by $k$ SNe is small. In fact, for constant models like Model A, the ratio $n_k/n_{k+1}$ is fixed, independently of the values of the input parameters.

\subsection{Other stochastic models}
\noindent
For reference below, we shall here briefly comment on two other stochastic
models of the chemical evolution of the Halo, that of 
Argast et al. (2000\nocite{argast00}) and that of Oey (2000\nocite{oey00}) 
and concentrate on important differences between the different approaches. 

\begin{figure}
 \resizebox{\hsize}{!}{\includegraphics{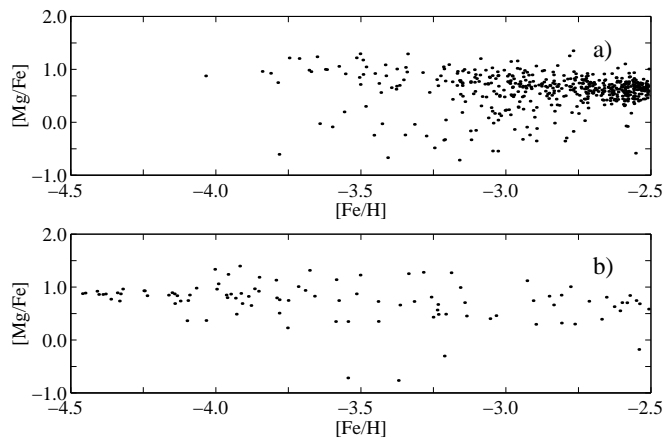}}
 \caption{A sample of model stars in the [Mg$/$Fe]--[Fe$/$H] plane. Stellar yields are taken from Nomoto et al. (1997). {\bf a)} Volume-limited sample of $500$ stars. {\bf b)} Biased sample (see Appendix \ref{app_renorm}) of $100$ stars.}
 \label{MgFeHNetal97_sc}
\end{figure}

\par

The model of Argast et al. (2000\nocite{argast00}) is formulated less in analytical terms
than ours, i.e. it is brought to numerical form more directly. In their model
the SFR scales in proportion to $\rho^2$, while our SFR is prescribed and, 
since space averages are used both for SFR and $\rho$, corresponds to a linear
density dependence in this particular model. Extra-galactic infall is not explicitly considered. 
Also, mixing is treated differently, with a mixing-mass of 
$5\times 10^4$ M$_\odot$ as a standard choice for each SN which first mixes 
its debris into that ISM mass before
any further star-formation may occur. Argast et al. (2000\nocite{argast00}) also mainly explored
the resulting abundances for low-mass stars for one set of yields. We shall
later make some comparisons between those results and ours.

\par

The model of Oey (2000\nocite{oey00}) is physically different in that it assumes that 
star-formation and supernova enrichment occurs in aggregates, i.e. giant 
star-forming regions and superbubbles. (We note, however, that just as her
model could be reformulated to the case of individual uncorrelated SNe, our
model could be reformulated to the aggregated case.) Technically, her model is 
an elegant inhomogeneous generalization of the classical one-zone models of 
chemical evolution. It is closed (i.e. no extra-galactic infall) and certain
uniformity assumptions must be made, concerning the filling factor of 
star-forming regions and the probability density
function for obtaining a certain metallicity at points affected by star
formation, both assumed to be the same for all stellar generations. Oey has
applied her basically static model to the study of metallicity distributions for stars
in the Galaxy (see also Oey 2003\nocite{oey03}), but she has not explored 
different abundance ratios more in detail.

\section{Results and discussion}\label{results}
\subsection{Terminology and methodology}
\noindent
In KG we introduced the terminology ``$A/$H diagrams''
for plots of abundance ratios of stars like [$A/B$] vs [$C/$H], where $A$, $B$ and $C$
denote different heavy elements and H, as usual, hydrogen. The square 
brackets have their standard meaning, denoting logarithmic abundance ratios
relative to the sun. Similarly, we denoted plots of
heavy element ratios like [$A/B$] vs [$C/D$], $D$ representing another heavy element, 
``$A/A$ diagrams''. We shall use this terminology below.

\begin{figure}
 \resizebox{\hsize}{!}{\includegraphics{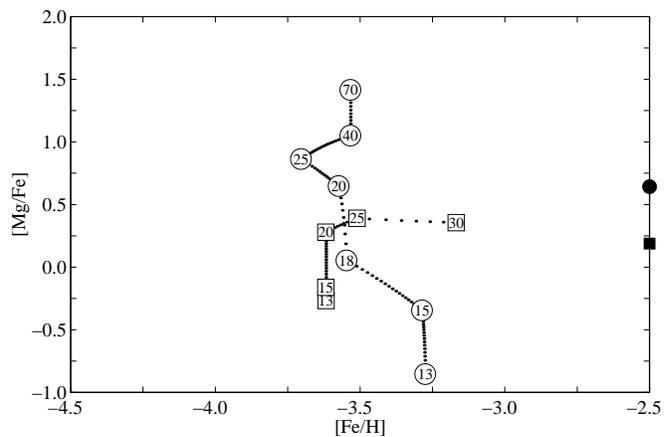}}
 \caption{Stellar yields of Mg and Fe in the range $13-70~\mathcal{M_{\odot}}$ (circles) by Nomoto et al. (1997) and $13-30~\mathcal{M_{\odot}}$ (squares) by UN02. The encircled numbers denote initial stellar mass. The dots connecting the symbols are distributed according to a Salpeter IMF. The filled circle and the square at the metal-rich end mark the IMF-averaged [Mg$/$Fe] ratio given the yields by Nomoto et al. (1997) and UN02, respectively. The [Fe/H] ratio is calculated given an average mixing mass of $\overline{M}_{\mathrm{mix}}=2.25\times 10^{5}~\mathcal{M_{\odot}}$.}
 \label{MgFeH_yields}
\end{figure}

\par

The synthetic $A/$H and $A/A$ diagrams will be presented both as probability density 
functions as well as scatter plots, where we have randomly picked a number of stars from 
the calculated density functions. Normally, these stars are picked with a metallicity 
distribution representative for the full Halo population of surviving low-mass stars, 
which means that the number of stars with very low metallicities is only a small
fraction of the total sample. However, in many surveys of Pop II stars one does
not study volume-limited samples, but deliberately aims at a biased
sample, e.g. such that a roughly equal number of stars is analysed for each
interval, of equal width, in logarithmic metallicity. Behind this choice may be
intentions to delineate abundance trends as functions of metallicity, or
a particular interest in the most metal-poor stars which requires biasing of
the survey towards the discovery of those objects. In our simulations we shall
show examples of both volume-limited samples of model stars, as well as 
samples biased towards equal number of stars for each bin in, e.g., [Fe$/$H].
The transformations from distributions of volume-limited samples to biased 
samples of stars is detailed in Appendix \ref{app_renorm}, below.

\begin{figure}
 \resizebox{\hsize}{!}{\includegraphics{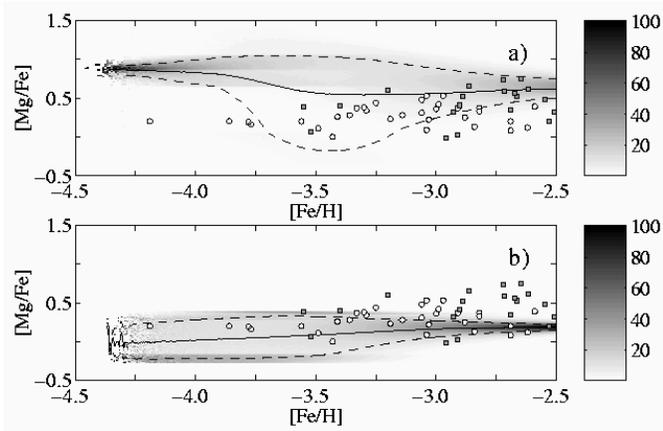}}
 \caption{Probability density functions of low-mass stars in the [Mg$/$Fe]--[Fe$/$H] plane. {\bf a)} The resulting density function using stellar yields by Nomoto et al. (1997). {\bf b)} The corresponding function using the yields by UN02. The density functions are renormalized to a biased sample (see Appendix B) and the shaded bars indicate the relative number density of stars in arbitrary units. The full line represents the mean [Mg$/$Fe] ratio at given [Fe$/$H] and the dashed lines indicate the scatter. The small circles denote extremely metal-poor giants observed by Cayrel et al. (2004) and the gray squares denote dwarfs observed by Cohen et al. (2004).}
 \label{MgFeH_df}
\end{figure}

\par

We shall now present a number of simulated $A/$H and $A/A$ diagrams, and in passing
compare them with some recent observations. We begin with a study of the $A/$H
diagrams of the so-called $\alpha$ elements Mg, Si, Ca and Ti and
defer a discussion of some particularities of the diagrams to the end of that study.

\subsection{A/H diagrams for $\alpha$ elements}\label{ahdiagrams}
\noindent
The different $\alpha$ elements Mg, Si, Ca and Ti are formed under different
conditions, and their abundances thus probe different phases of the 
evolution of massive stars towards the supernova stage. Mg originates from
the hydrostatically processed carbon core, and from explosive Ne-C burning.
The amount of Mg produced (as well as of O and Ne) varies considerably with
stellar initial mass, and with the extent of mixing. Si and Ca originate
from explosive O- and Si-burning  and their yields vary less with initial mass.
They probe the progenitor model but also depend on the energy in the SN 
explosion and the amount of fall-back. The energy determines the excitation
of the shock front passing through the model, which in turn determines the 
alpha-rich freeze-out from explosive Si burning, and thus the Ti yields. 

\par

{\it (a) Mg/Fe}. As a first example we show in Fig. \ref{MgFeHNetal97_sc}a the distribution 
of a volume-limited 
sample of $500$ model stars in the [Mg$/$Fe]-[Fe$/$H] diagram. Here, we have used
the SN yields of Nomoto et al. (1997\nocite{netal97}), essentially the same yields as used by 
Argast et al. (2000\nocite{argast00}). As seen when comparing to their Fig. 2, the two 
plots give a similar impression, which is to be expected in view of the relatively similar
chemical evolution models. They both show a [Mg$/$Fe] value of $0.5$ around [Fe$/$H]$ 
= -2.5$, with a minority population of stars with smaller Mg abundances, and an 
increasing scatter for [Fe$/$H] decreasing below $-3.0$. A difference is that 
Argast et al. (2000\nocite{argast00}) find stars 
even more Mg-poor than our extremes, by about a factor of two.
The result is, as discussed by Argast et al. (2000\nocite{argast00}), not in accordance
with observations, which do not disclose stars with [Mg$/$Fe] values 
significantly below $0.0$. We also note that the more recent surveys of 
abundances in stars of Extreme Pop II, Cayrel et al. (2004\nocite{cayrel04}) and 
Cohen et al. (2004\nocite{cohen04}), do not show a single star with a negative [Mg$/$Fe] ratio. 
We have also used the more recent yield calculations of UN02\nocite{un02}. 
Their Mg yields are systematically greater, relative to those of Fe, by a factor of $3$ for the low
initial masses, while they are smaller (but with Mg$/$Fe still greater than 
solar) for the high-mass stars (see Fig. \ref{MgFeH_yields}). 
The mean value of [Mg$/$Fe] in the metal-rich end is now only $0.2$ dex, which departs from 
the observed value, while the most metal-poor model stars are now closer to 
observations although some of them still too Mg-poor.

\begin{figure}
 \resizebox{\hsize}{!}{\includegraphics{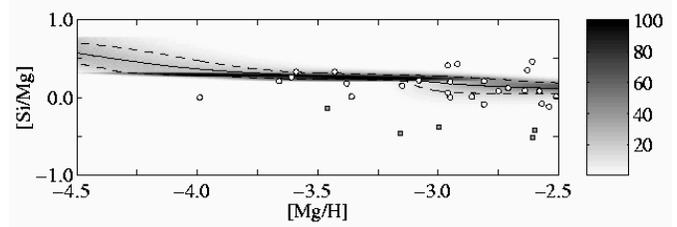}}
 \caption{Probability density function of low-mass stars in the [Si$/$Mg]--[Mg$/$H] plane. Stellar yields are taken from WW95. Symbols, see \mbox{Fig. \ref{MgFeH_df}.}}
 \label{SiMgH_df}
\end{figure}

\begin{figure}
 \resizebox{\hsize}{!}{\includegraphics{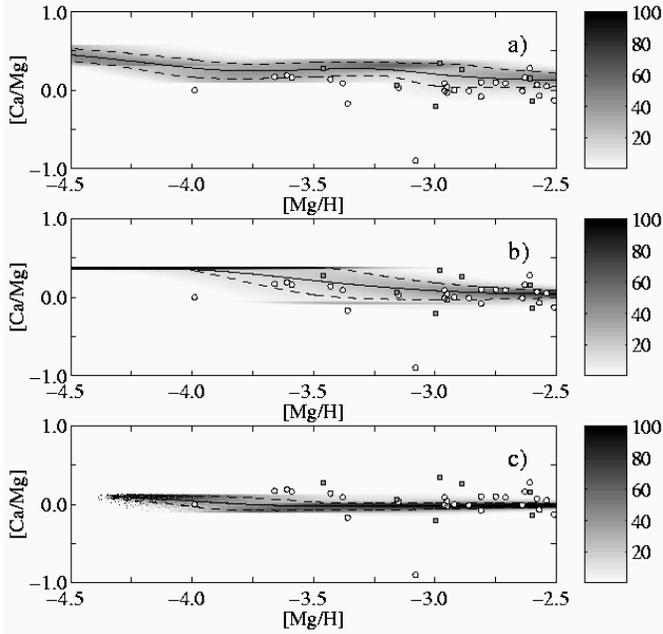}}
 \caption{Probability density function of low-mass stars in the [Ca$/$Mg]--[Mg$/$H] plane. {\bf a)} The resulting distribution of stars if yields from WW95 are used. {\bf b)} The corresponding distribution using yields by UN02. {\bf c)} Yields by CL04. Symbols, see Fig. \ref{MgFeH_df}.}
 \label{CaMgH_df}
\end{figure}

\begin{figure}
 \resizebox{\hsize}{!}{\includegraphics{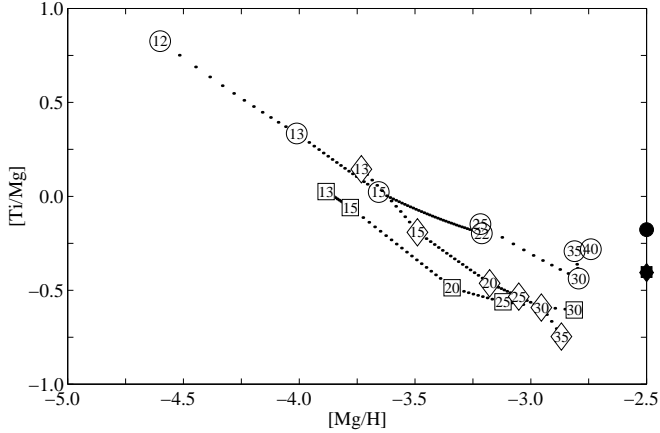}}
 \caption{Three different sets of stellar yields of Ti and Mg. Yields from WW95 are indicated by circles ($12-40~\mathcal{M_{\odot}}$) while yields from UN02 are indicated by squares ($13-30~\mathcal{M_{\odot}}$) and  yields from CL04 by diamonds ($13-35~\mathcal{M_{\odot}}$). Otherwise, the notation is as in Fig. \ref{MgFeH_yields}.}
 \label{TiMgH_yields}
\end{figure}

\par

In Fig. \ref{MgFeHNetal97_sc}b we display a sample of $100$ model stars, again 
using the yields by Nomoto et al. (1997), biased to equal number for equal 
intervals in [Fe$/$H]. We note here that, unexpectedly, the scatter in 
[Mg$/$Fe] diminishes for the most metal-poor stars. That is, there seems to be
a departure from the general law often taken for granted (e.g. by Nissen
et al. 1994\nocite{netal94}) and demonstrated by the simulations of Argast et al. (2000\nocite{argast00})
for a number of element ratios: that the scatter should increase towards the 
lowest metallicities as a reflection of the small number $k$ of SNe (essentially
a normal stochastic $k^{-1/2}$ dependence). Here, instead we find a
diminishing scatter in the most metal-poor end of the diagram. 
To study the reason for this we calculated the density function in detail
(cf. Appendix \ref{app_derivation}), biased in equal [Fe$/$H] intervals. The resulting distribution
and scatter is shown in Fig. \ref{MgFeH_df}a. A closer look at the yields used 
(Nomoto et al. 1997\nocite{netal97}), shows that the lowest Fe yields are obtained
for SNe with initially $25$ solar masses, and these give a [Mg$/$Fe] 
value of $0.8$ (see Fig. \ref{MgFeH_yields}), as observed for the most 
metal-poor stars in Figs. \ref{MgFeHNetal97_sc}b 
and \ref{MgFeH_df}a. Since the selection of contributing SNe for stars with the 
lowest iron abundances will be heavily biased towards those that contribute the lowest
iron yields, the distribution in the diagram for the lowest metallicities will
narrow down in [Mg$/$Fe]. Recall that stars with such a low metallicity were 
formed in the largest mixing volumes in which the yields were diluted 
with the largest mixing masses from the massive tail of the mixing mass 
distribution $f_{M_{k=1}}$, i.e., larger than 
\mbox{$\overline{M}_{\mathrm{mix}}=2.25\times 10^{5}~\mathcal{M_{\odot}}$} which 
corresponds to the average mixing mass $\langle M_{\mathrm{mix}}\rangle$,
as used for calculating the ``standard'' metallicity in, e.g., Fig. \ref{MgFeH_yields}. 

\par

In Fig. \ref{MgFeH_df}b we display the analogous distribution,
however calculated with the yields from UN02\nocite{un02}. Here, the 
situation is different, mainly because the supernova masses with the lowest
Fe yields include both masses with the lowest Mg$/$Fe ratios, [Mg$/$Fe]$ = -0.27$,
{\it and} the slightly higher masses with much higher Mg yields, [Mg$/$Fe]$ = 0.28$, as 
shown in Fig. \ref{MgFeH_yields}. Individual SNe of both masses may contribute to 
the Mg$/$Fe ratio for stars with [Fe/H] around $-4$. 

\begin{figure}
 \resizebox{\hsize}{!}{\includegraphics{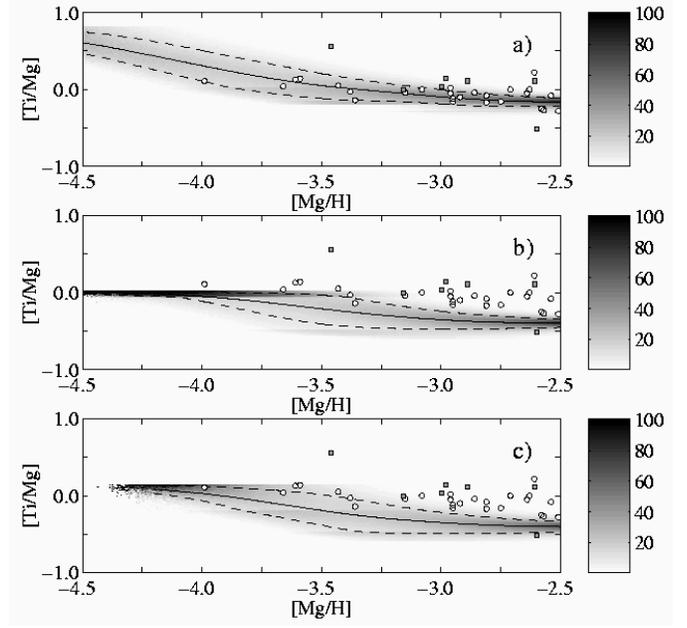}}
 \caption{Probability density functions of low-mass stars in the [Ti$/$Mg]--[Mg$/$H] plane. {\bf a)} Yields by WW95. {\bf b)} Yields by UN02. {\bf c)} Yields by CL04. Symbols, see Fig. \ref{MgFeH_df}.}
 \label{TiMgH_df}
\end{figure}

\par

Arnone et al. (2005\nocite{arnone04}) have recently explored the scatter in 
[Mg$/$Fe] for $23$ turn-off stars with 
$-3.4 \le \mathrm{[Fe}/\mathrm{H]}\le -2.2$, and found an rms scatter of 
$0.06$, no larger than what is expected from uncertainties in the analysis.
The different sets of yields used in this study all predict significantly
greater scatter in [Mg$/$Fe] when they are applied to our inhomogeneous 
model. This will be commented on below.

\par

A short comment should be made on the definition of the mean and scatter indicated by the full and 
dashed lines, respectively, in, e.g., Fig. \ref{MgFeH_df}. The mean is the usual average given by 
$\langle$[Mg$/$Fe]$\rangle$, at given [Fe$/$H]. The scatter is defined such that the 
fraction of stars outside the dashed lines is $32\%$ (at given [Fe$/$H]), $16\%$ at each side.

\par  

\begin{figure}[t]
 \resizebox{\hsize}{!}{\includegraphics{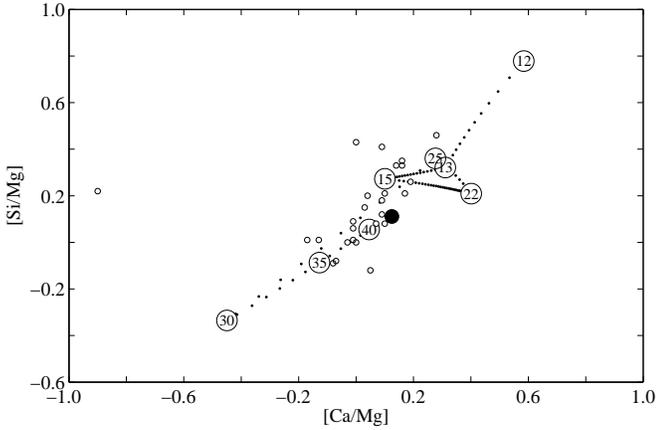}}
 \caption{The yields by WW95 plotted in the $[\mathrm{Si}/\mathrm{Mg}]-[\mathrm{Ca}/\mathrm{Mg}]$ plane. The notation is as in Fig. \ref{MgFeH_yields}. The big black dot marks the IMF-averaged point. The extremely metal-poor giants ($[\mathrm{Mg}/\mathrm{H}]<-2.50$) observed by Cayrel et al. (2004) are denoted by small circles.}
 \label{SiCaMg_yields}
\end{figure}

{\it (b) Si/Mg}. In Fig. \ref{SiMgH_df} we show the density function of low-mass stars, biased to equal 
distribution in [Mg$/$H], in the [Si$/$Mg]--[Mg$/$H] diagram. The yields by 
WW95\nocite{ww95} were used here; somewhat positive
[Si$/$Mg] values also show up when the yields of UN02\nocite{un02} or
CL04\nocite{cl04} are used. This moderate excess of Si relative to Mg 
(as compared with the solar ratio) is found observationally by Cayrel et al. (2004\nocite{cayrel04}), 
while the observations for dwarfs by Cohen et al. (2004\nocite{cohen04}) indicate significantly lower 
Si$/$Mg ratios. The sloping tendency in the [Si$/$Mg]--[Mg$/$H] diagram, which  
may also be traced if UN02\nocite{un02} yields are used, is not found in 
the observed values. The origin of the slope is the variation of the Si$/$Mg ratio
with the Mg yield -- the lowest SN masses, contributing least Mg and thus 
leading to stars in the left in the diagram,
have the highest Si$/$Mg ratio, then the ratio levels off at $0.3$ dex for a range 
of masses (from $13~\mathcal{M_{\odot}}$ to $22~\mathcal{M_{\odot}}$) with higher Mg yields 
and finally drops to even lower ratios (cf. Fig. \ref{SiCaMg_yields}). A similar, though less dramatic, 
pattern is shown by the UN02\nocite{un02} yields. The relatively constant 
level for intermediate SN masses in the WW95 yields also leads to the minimum 
in the dispersion in [Si$/$Mg] around [Mg$/$H]$ = -3.5$ in Fig. \ref{SiMgH_df}.

\par

{\it (c) Ca/Mg}. The scatter in [Ca$/$Mg] as a function of [Mg$/$H], 
displayed in Fig. \ref{CaMgH_df}b (i.e., UN02), shows the 
opposite behavior to that of [Si$/$Mg] in Fig. \ref{SiMgH_df}. For the dispersion in
[Ca$/$Mg] there is a maximum, instead of a minimum, around [Mg$/$H]$ = -3.5$. 
We also see a decreasing trend of [Ca$/$Mg] with increasing [Mg$/$H]. The yields by UN02
 show a strong decrease of the Ca$/$Mg ratio as the Mg yields (and SN mass) increase. 
When WW95\nocite{ww95} yields, with a similar but more irregular behavior, are used
instead, the dispersion stays more constant as a function of [Mg$/$H] while the 
sloping trend in the diagram still prevails. Both sets of yields, however,
suggest significantly positive [Ca$/$Mg] values for the most metal-poor stars,
a tendency which is found neither in the Cayrel et al. (2004\nocite{cayrel04}), 
nor the Cohen et al. (2004) observed values.
For the CL04\nocite{cl04} yields, the scatter stays small (reflecting
a small mass-dependence of the yields), the slope only occurs for the stars with the very 
smallest [Mg$/$H] values, and the [Ca$/$Mg] values remain about solar, as observed.

\begin{figure}
 \resizebox{\hsize}{!}{\includegraphics{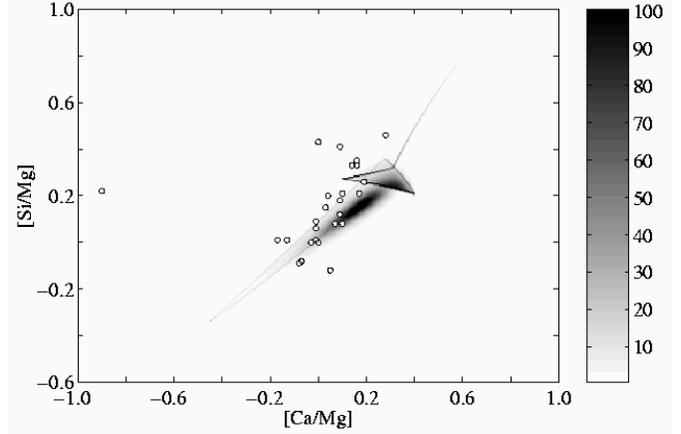}}
 \caption{The full density function of stars in the $[\mathrm{Si}/\mathrm{Mg}]-[\mathrm{Ca}/\mathrm{Mg}]$ plane (cf. Fig. \ref{SiCaMg_yields}) plotted together with the observations by Cayrel et al. (2004).}
 \label{SiCaMgWW95_df}
\end{figure}

\begin{figure}[t]
 \resizebox{\hsize}{!}{\includegraphics{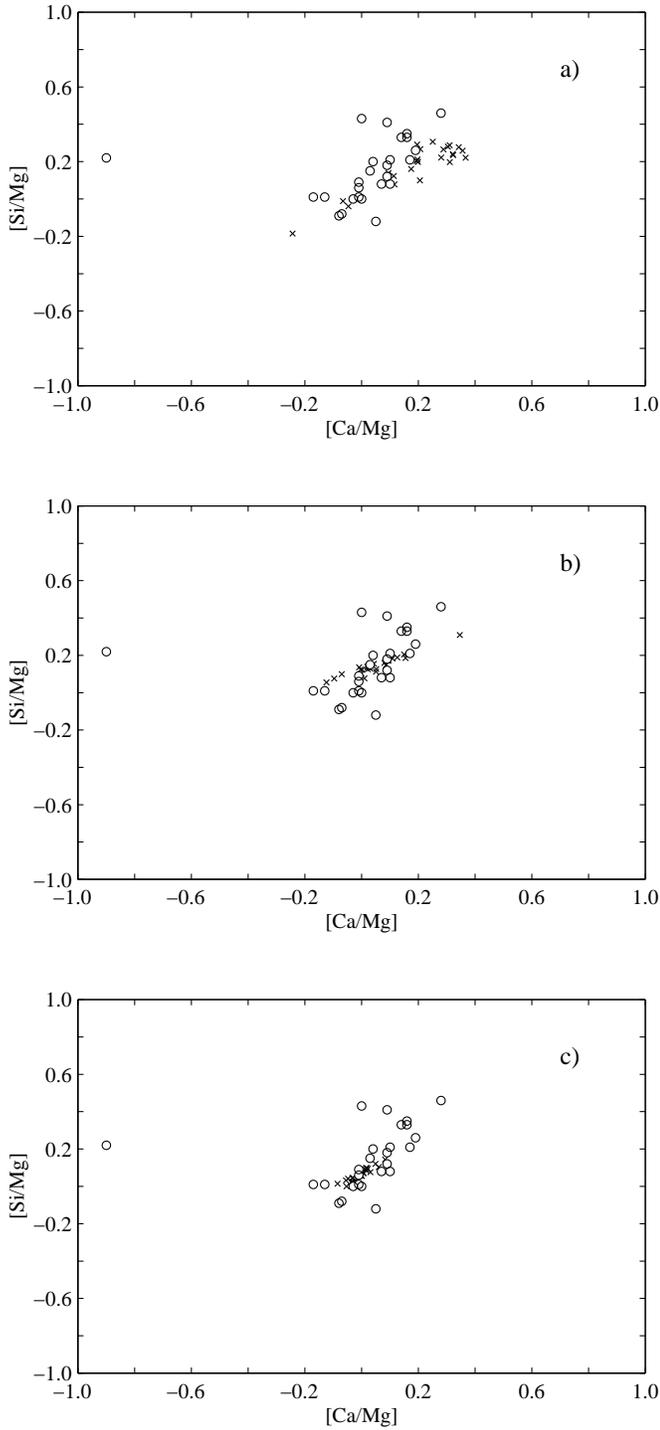}}
 \caption{A sample of $25$ model stars (crosses) plotted together with the observations by Cayrel et al. (2004, small circles). {\bf a)} The model stars are drawn from a density function where we used stellar yields by WW95. {\bf b)} Yields by UN02. {\bf c)} Yields by CL04.}
 \label{SiCaMg_sc}
\end{figure}

\par

{\it (d) Ti/Mg}. The relative yields of Ti$/$Mg relative to Mg for different SN initial masses
are displayed in Fig. \ref{TiMgH_yields}, for the three different sets of yields. Here, a clear
negative slope in Ti$/$Mg with increasing mass, and increasing Mg yield, is seen
for all three sets, although the WW95\nocite{ww95} Ti yields are 
systematically higher. It is easy to predict that this slope should lead 
to a decreasing trend in [Ti$/$Mg] with increasing [Mg$/$H], for the most 
metal-poor stars. The model simulations also clearly display such a trend as is seen in
Fig. \ref{TiMgH_df}. Although, some weak trend of that character may possibly be traced 
in the data of Cayrel et al. (2004\nocite{cayrel04}), as well as in Cohen et al. (2004\nocite{cohen04}), 
it is not at all as pronounced as in the simulations. We also note that the high [Ti$/$Mg] 
ratios predicted by the WW95 yields for the most metal-poor stars 
are not yet observed; obviously, the two other sets of yields on the other hand seem to
give too low [Ti$/$Mg] ratios.

\par 

{\it (e) Conclusions from the $A/$H diagrams of the $\alpha$ elements}.
The difference between the $A/$H diagrams calculated with different sets of yields as
discussed above clearly show the possibilities of the simulations,
when compared with observations of the most metal-poor stars, to illuminate the 
adequacy and the short-comings of different yield calculations. Within the assumptions of 
the model, the simulations also
suggest an explanation for the astonishingly small scatter that appears in 
observed abundance ratios even for the most metal-poor stars. This could be
the result of cosmic selection effects due to the (trivial) fact that the most 
metal-poor stars, formed of gas only polluted by one single SN, may predominantly 
originate in such clouds that are affected by SNe that produce a minimum of pollution.
However, except for the unexpected small scatter in abundance ratios, 
there is no support for that hypothesis in the comparison
with observations. The shift from one particular mass of the SNe to
another one, as the metallicity increases, leads to predicted trends with
metallicity that are generally not observed. Also, the narrow horizontal stripes in
the left region of some $A/$H diagrams (see, e.g., Figs. \ref{MgFeH_df}a and \ref{CaMgH_df}b), 
mapping the effect of one dominating SN mass in the simulations, are not observed as yet. 

\begin{figure}[t]
 \resizebox{\hsize}{!}{\includegraphics{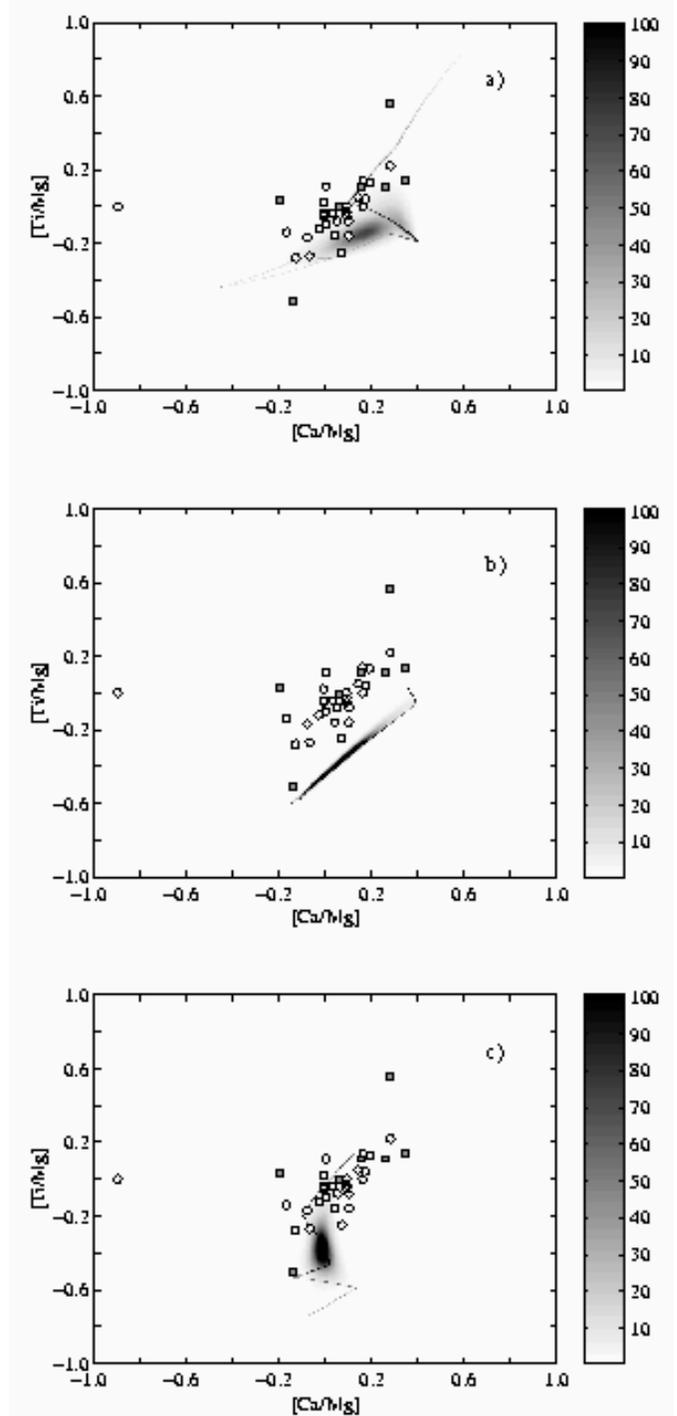}}
 \caption{Distribution of low-mass stars in the [Ti$/$Mg]--[Ca$/$Mg] plane. Small circles denote the observations of extremely metal-poor giants ([Mg$/$H]$<-2.5$) by Cayrel et al. (2004), and the gray squares denote a corresponding sample ([Mg$/$H]$<-2.5$) of dwarfs by Cohen et al. (2004). {\bf a)} Yields by WW95. {\bf b)} Yields by UN02. {\bf c)} Yields by CL04.}
 \label{TiCaMg_df}
\end{figure}

\par

Several different hypotheses may be thought of as explanations for this 
short-coming of the model. One may be that the mixing masses are considerably
larger than expected (at least a factor of ten), which will shift all 
the model low-mass stars towards lower metallicities. If so, we should
be able to find a continuous distribution of stars down to [Fe$/$H]$< -5$
(see Paper~I). Possibly, the mixing mass distributions for high numbers 
$k$ of contributing SNe could be significantly shifted towards higher 
mixing masses, e.g., due to a large increase in the infall rate of pristine 
gas. This will shift a number of model stars from the right (higher 
metallicity) side in the $A/$H diagrams to the left side (lower metallicity). 
Since the stellar distribution is heavily skewed towards the more metal-rich 
side, such a shift will lead to a considerably reduced stochastic scatter 
(i.e., a $k^{-1/2}$ effect). Another related possibility is that mixing with
metal-free gas also occurs due to stochastic infall of intergalactic clouds. 
If this mechanism is to be efficient, it will require a considerable infall 
of clouds with masses on the order of $10^6~\mathcal{M_{\odot}}$ at a rate 
of several clouds per 1 million years, and a significant correlation between 
infall of these clouds and star formation; conditions that cannot be ruled
out. A third interesting possibility would be that the dominating population
of the most metal-poor stars in fact are Pop III stars, with originally 
no or very small heavy-element abundances, but polluted by 
gas from several SNe, accreted later, for e.g. when the stars
passed through the Galactic disk (see Yoshii 1981\nocite{yoshii81}; 
Yoshii et al. 1995\nocite{ymk95}; Shigeyama et al. 2003\nocite{sty03}, see 
also the discussion in Christlieb et al. 2004\nocite{cetal04}). The small 
scatter in abundance ratios would then just reflect the well-mixed 
interstellar gas at later Galactic epochs. The possibility that this is 
the case is dependent on a number of uncertain conditions, such as the 
distribution and inhomogeneities of interstellar matter and the degree 
to which the stars are shielded from accretion by their own stellar winds.

\par

Arnone et al. (2005\nocite{arnone04}) have recently discussed their 
observation that the inhomogeneous models of Argast et al. (2000) predict 
a far greater scatter in [Mg$/$Fe] than observed 
(cf. also Argast et al. 2002\nocite{argast02}). In their detailed
discussion, Arnone et al. (2005\nocite{arnone04}) suggest the possibility 
that shorter mixing time-scales, or longer cooling time scales than used 
in the inhomogeneous models could be the explanation. In the latter 
case, the suggestion is that longer cooling times before next generation 
of low-mass stars forms would admit many SNe of different masses to 
explode and contribute to the gas.

\begin{figure*}[t]
 \resizebox{\hsize}{!}{\includegraphics{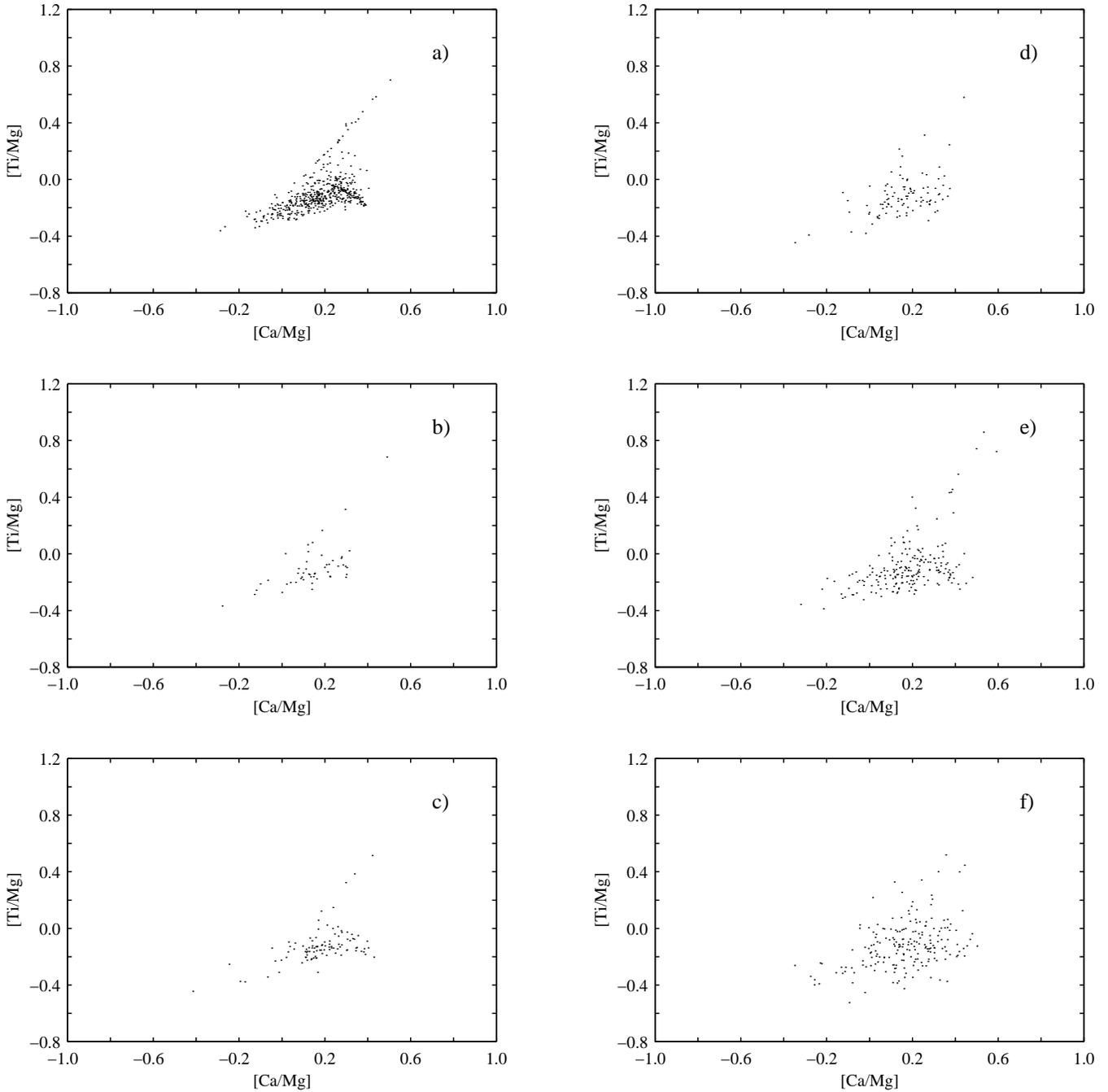}}
 \caption{Model stars in the [Ti/Mg]--[Ca/Mg] plane drawn from the density function displayed in Fig. \ref{TiCaMg_df}a. {\bf a)} 500 stars with no uncertainty in the abundance ratios. {\bf b)} 50 stars with no uncertainty. {\bf c)} 100 stars with an uncertainty of $0.02$ dex applied in both abundance ratios. {\bf d)} 100 stars with an uncertainty of $0.05$ dex. {\bf e)} 200 stars with an uncertainty of $0.05$ dex. {\bf f)} 200 stars with an uncertainty of $0.10$ dex.}
 \label{TiCaMg_scatter}
\end{figure*}

\par

Whatever the reason for the tendencies for mismatch between the 
simulated and observed $A/$H diagrams may be, we shall now proceed to
$A/A$ diagrams which, as they are much less dependent on assumptions
concerning mixing masses and infall of pristine gas than the $A/$H
diagrams, should give more robust results.

\subsection{$A/A$ diagrams for $\alpha$ elements}
\noindent
In Fig. \ref{SiCaMg_yields} we have plotted the Si$/$Mg versus the Ca$/$Mg ratios 
of the yields of WW95 and in Fig. \ref{SiCaMgWW95_df} the corresponding 
predicted distribution of low-mass stars. In both figures the observations by 
Cayrel et al. (2004) are also plotted. In general, fairly good agreement is 
found. We note, however, the 
outlier at [Ca$/$Mg]$=-0.90$ (BS 16477-003) which cannot be explained by the yields of 
WW95 or any of the other sets of yields discussed here.
As discussed in KG, the distribution of low-mass stars in the $A/A$ diagrams is mainly
determined by the stellar yields.  In fact, this is illustrated by comparing 
Fig. \ref{SiCaMg_yields} and Fig. \ref{SiCaMgWW95_df}. Visually, the two $A/A$ diagrams appear 
very similar and the sequence of yield ratios in Fig. \ref{SiCaMg_yields} directly reveals the 
location of the ``spurs'' in Fig. \ref{SiCaMgWW95_df} (see Sect. \ref{spurs}, below), although 
none of the ``spurs'' reaches the outlier.

\par

Alternatively, we have in Fig. \ref{SiCaMg_sc} plotted $25$ model stars in the [Si$/$Mg]-[Ca$/$Mg] 
diagram, drawn from the calculated density functions resulting from the different sets of yields 
(cf. Fig. \ref{SiCaMgWW95_df} and Fig. \ref{SiCaMg_sc}a). For all sets, there is nice general 
agreement with the observations of Cayrel et al. (2004\nocite{cayrel04}). The observations by 
Cohen et al. (2004) have been excluded from these diagrams.
It is also noteworthy that the general trend, suggested by the models, of a positive
correlation of Si and Ca excesses relative to Mg is verified by the 
observations, at least qualitatively. Regarding the scatter in the
observations, as compared with the sequences delineated by the models, it seems
that most of it may be explained as a result of observational uncertainties.

\par

Similarly, in Fig. \ref{TiCaMg_df}  the [Ti$/$Mg]-[Ca$/$Mg] diagrams are displayed. Here, the 
synthetic distributions of stars are in the form of probability density functions. There is 
a tendency for the models to predict somewhat low Ti abundances, as compared
with the observations of dwarfs as well as of giants. The yield set of
UN02\nocite{un02} results in a narrow sequence in the diagram, with a 
slope close to the observed one but with an off-set by about a factor of $2$. 
Shifting the model stars by this off-set and introducing a Gaussian scatter of $\sigma = 0.10$ 
in both coordinates, a reasonable estimate of the observational errors, the resulting 
distribution agrees well with the observed one. We may conclude that the real cosmic scatter 
in this diagram around the model sequence may well be very small.

\subsection{The existence of ``spurs'' or ``SSSs'' in the $A/A$ diagrams}\label{spurs}
\noindent
The extended narrow sequences, ``spurs'', in the calculated distributions (see, e.g., 
Fig. \ref{TiCaMg_df}a) represent the location of stars with major contributions 
from just a single supernova, though with different SN masses
(see KG for more discussion as well as Idiart \& Thevenin 2000\nocite{it00} for a first, 
not fully convincing, attempt to trace such features). We will call these
features Single Supernova Sequences (SSSs) here.  An interesting issue
is whether such sequences may really be traced observationally. In order to
study that, we have explored the distribution of model stars,
as the number of stars $N_{\star}$, as well as the random errors in [Ti$/$Mg] and [Ca$/$Mg], 
are varied. We use the yields of WW95\nocite{ww95}. Some 
resulting diagrams are presented in Fig \ref{TiCaMg_scatter}. From this we judge that observations 
of at least $200$ stars are needed for tracing an SSS,
if statistical errors in the abundance observations are about $0.05$ dex 
(Fig. \ref{TiCaMg_scatter}e) -- if the errors, e.g. in a differential analysis, are 
reduced to $0.02$ dex the SSSs in the diagrams could appear at $N_{\star}\lesssim 100$ 
(cf. Fig. \ref{TiCaMg_scatter}c), depending on the 
details in the shapes of these sequences.
Contemporary differential analyses of stars of this type may possibly
reach this differential accuracy (cf. Gustafsson 2004\nocite{G04}), but 
this is not the normal case. Also, 
for the purpose of finding the SSSs, it is important to
have reliable estimates of the random error in the abundances, since the sharp 
SSSs may then be found by deconvolution. Some guidance from expectation built 
on theoretical yields may perhaps also be useful. It is obviously
a difficult but interesting challenge to trace such features by going to large
samples and accurate observations of extreme Pop II stars. If found, they will
give direct evidence concerning supernova yields as well as upper IMFs for
the first generation of stars. Note that a failure of finding them does
not necessarily indicate that mixing between different supernova remnants
have occurred before the presently observed low-mass stars formed -- another
possible explanation for such a failure might be the existence of another 
parameter besides initial mass determining the SN II yields, such as initial 
angular momentum.

\begin{figure}
 \resizebox{\hsize}{!}{\includegraphics{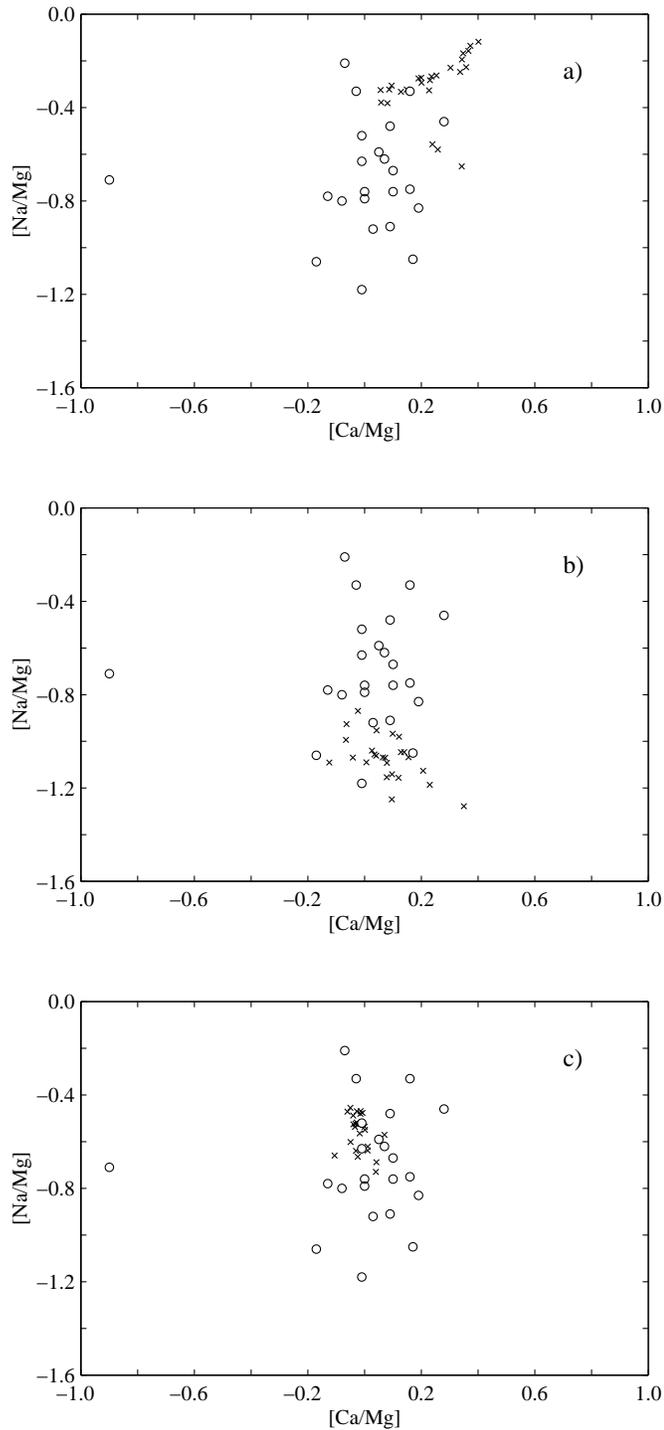}}
 \caption{[Na$/$Mg] vs. [Ca$/$Mg] for a sample of $25$ model stars (crosses). Symbols, see Fig. \ref{TiCaMg_df}. The model stars are drawn from density functions calculated using stellar yields by {\bf a)} WW95, {\bf b)} UN02, and {\bf c)} CL04.}
 \label{NaCaMg_sc}
\end{figure}

\begin{figure}
 \resizebox{\hsize}{!}{\includegraphics{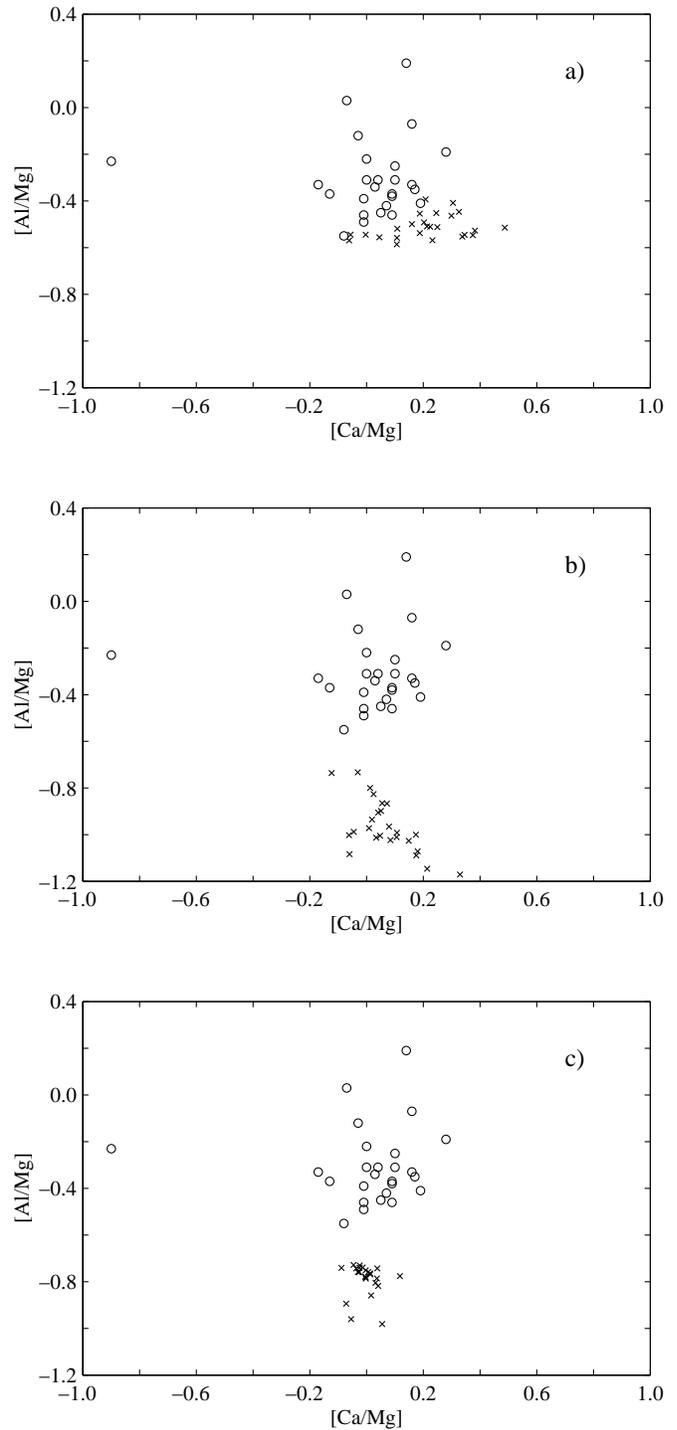}}
 \caption{[Al$/$Mg] vs. [Ca$/$Mg] for a sample of $25$ model stars (crosses). Symbols, see Fig. \ref{TiCaMg_df}. The model stars are drawn from density functions calculated using stellar yields by {\bf a)} WW95, {\bf b)} UN02, and {\bf c)} CL04.}
 \label{AlCaMg_sc}
\end{figure}

\subsection{$A/A$ diagrams for Na and Al}
\noindent
The elements Na and Al, both with an odd number of nucleons, are sensitive
to various model uncertainties in the calculation of yields. The strengths of 
abundance criteria for these elements are also subject to effects of
departures from LTE, which are in themselves uncertain. 

\begin{figure}
 \resizebox{\hsize}{!}{\includegraphics{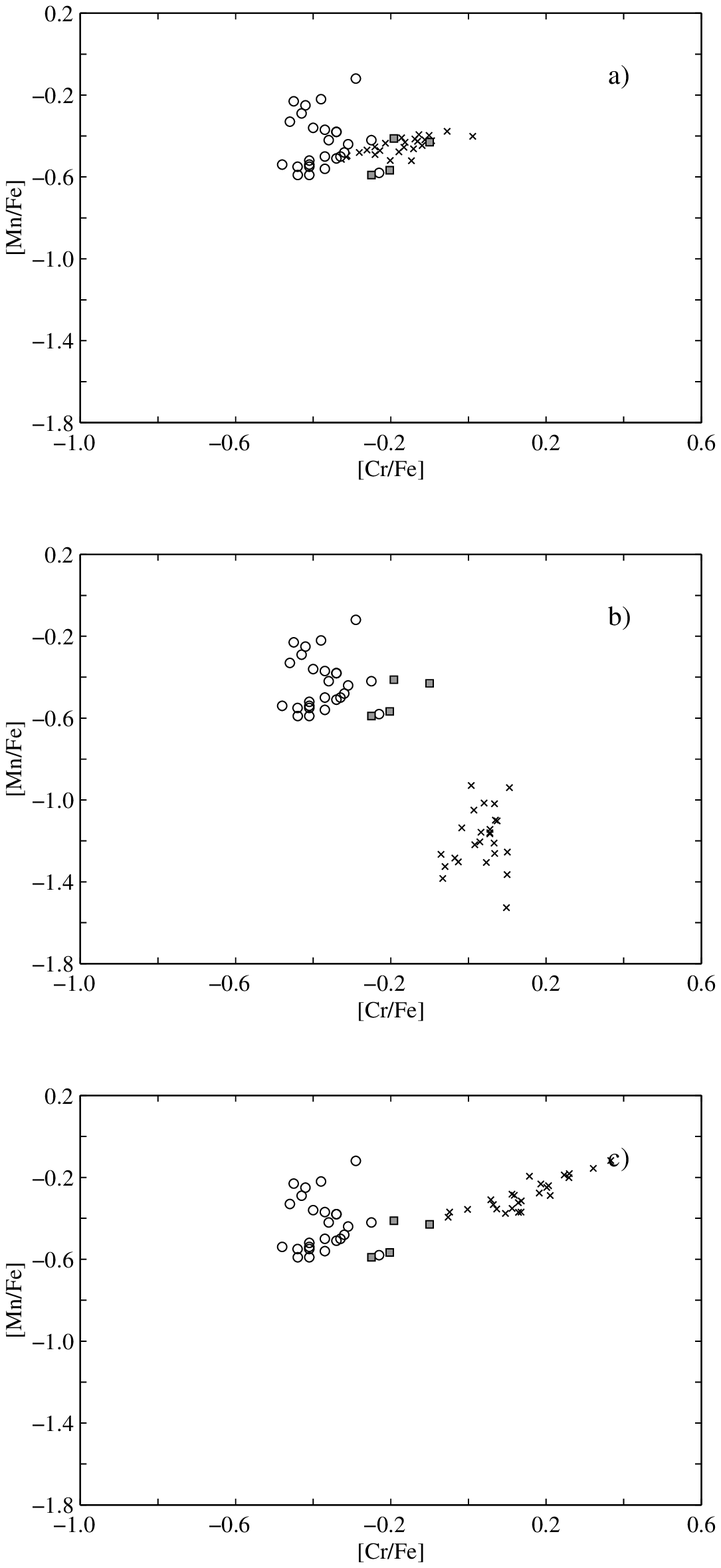}}
 \caption{[Mn$/$Fe] vs. [Cr$/$Fe] for a sample of $25$ model stars (crosses). Symbols, see Fig. \ref{TiCaMg_df}. The model stars are drawn from density functions calculated using stellar yields by {\bf a)} WW95, {\bf b)} UN02, and {\bf c)} CL04.}
 \label{MnCrFe_sc}
\end{figure}

\par

In Fig. \ref{NaCaMg_sc}, we display a sample of $25$ model stars in the 
[Na$/$Mg]-[Ca$/$Mg] diagram, using three different sets of yields. As usual, the 
open circles represent observed abundance ratios of Cayrel et al. (2004\nocite{cayrel04}), 
where the Na abundances are systematically corrected by $-0.5$ dex for non-LTE effects  
according to these authors, following Baum\"{u}ller et al. (1998\nocite{bbg98}). It is seen 
that the three sets of yields depart in magnitude by as much as $0.7$ dex 
(such that they correspond to the observed upper level, lower level, 
and approximate mean level of [Na$/$Mg], respectively).
The large observed scatter in [Na/Mg] is not reproduced by any of the 
sets of yields. We conclude that either the spurious 
errors in abundances (including the consideration of 
non-LTE effects) at least are on the order of  0.25 dex, or that the 
scatter in real SN yields are greater than predicted by contemporary models. A third
possibility is that the abundances of some of these very metal-poor giants 
might be affected by hot proton burning in the ON, NeNa and MgAl cycles, as 
has been found for stars in globular clusters, c.f., e.g., 
Gratton et al. (2001\nocite{gratton01}). These authors find that internal mixing 
mechanisms in stars are not sufficient to explain 
the abundance effects in globular clusters but suggest mass-loss
from intermediate-mass AGB stars to be active, either by contributing to
lower-mass stars in slow star-forming regions (Cottrell \& Da Costa 1981\nocite{cd81})
or by polluting the surface layers of low-mass stars (D'Antona et al. 1983\nocite{dantona83}). 
If any of these explanations would be valid, many of the most metal poor stars 
must have been formed in very dense and rather long-lived star-forming regions.

\begin{figure}
 \resizebox{\hsize}{!}{\includegraphics{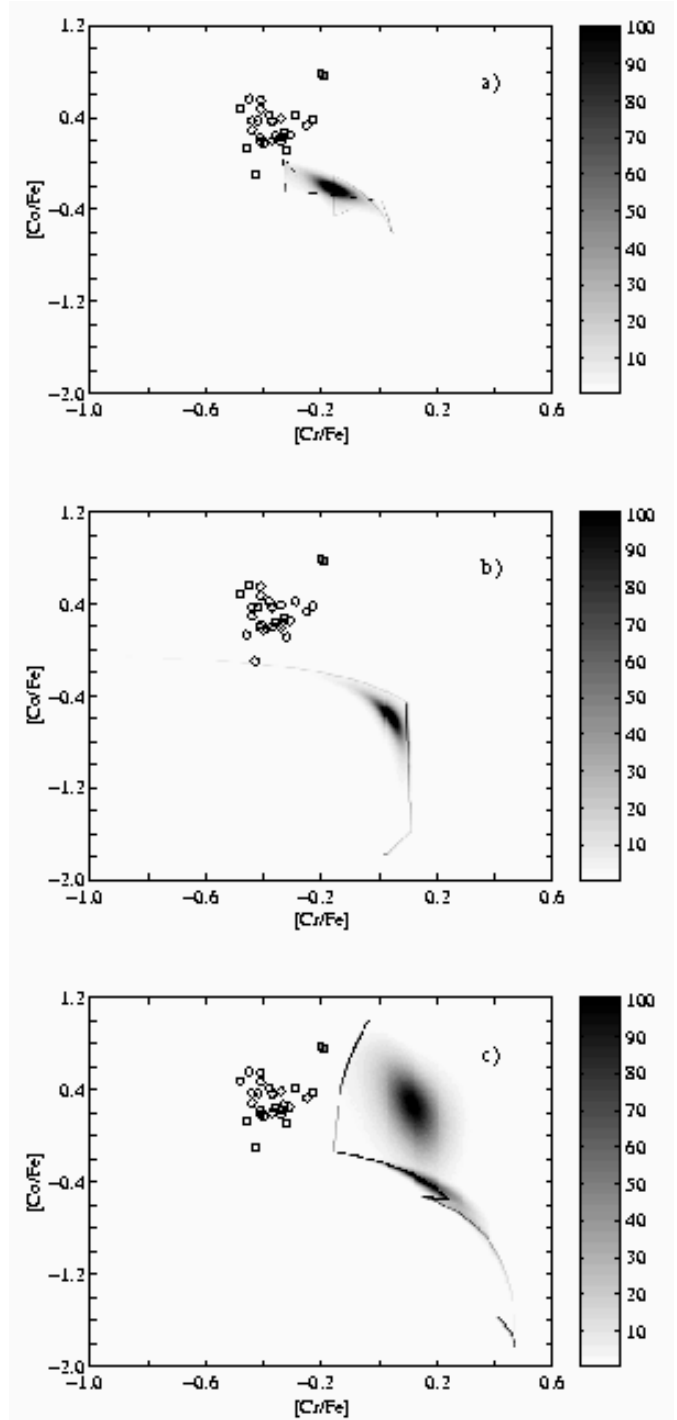}}
 \caption{Density functions of low-mass stars in the [Co$/$Fe]--[Cr/Fe] plane. Symbols, see Fig. \ref{TiCaMg_df}. {\bf a)} Yields by WW95. {\bf b)} Yields by UN02. {\bf c)} Yields by CL04. The density function is multiplied by a factor of $5$ in order to enhance the contrast.}
 \label{CoCrFe_df}
\end{figure}

\par

Some further support in this direction is obtained from the [Al$/$Mg]-[Ca$/$Mg]
diagrams, Fig. \ref{AlCaMg_sc}. Here, the observed Al abundances have been adjusted upwards 
by $0.65$ dex to correct for non-LTE effects, according to Cayrel et al. (2004\nocite{cayrel04}),
following Baum\"{u}ller $\&$ Gehren (1997\nocite{bg97}) and Norris et al. (2001\nocite{nrb01}).
We find the [Al$/$Mg] ratios as calculated from the yields to be systematically
too low by $0.2 - 0.5$ dex, and with a predicted scatter less than the observed one.

\begin{figure}
 \resizebox{\hsize}{!}{\includegraphics{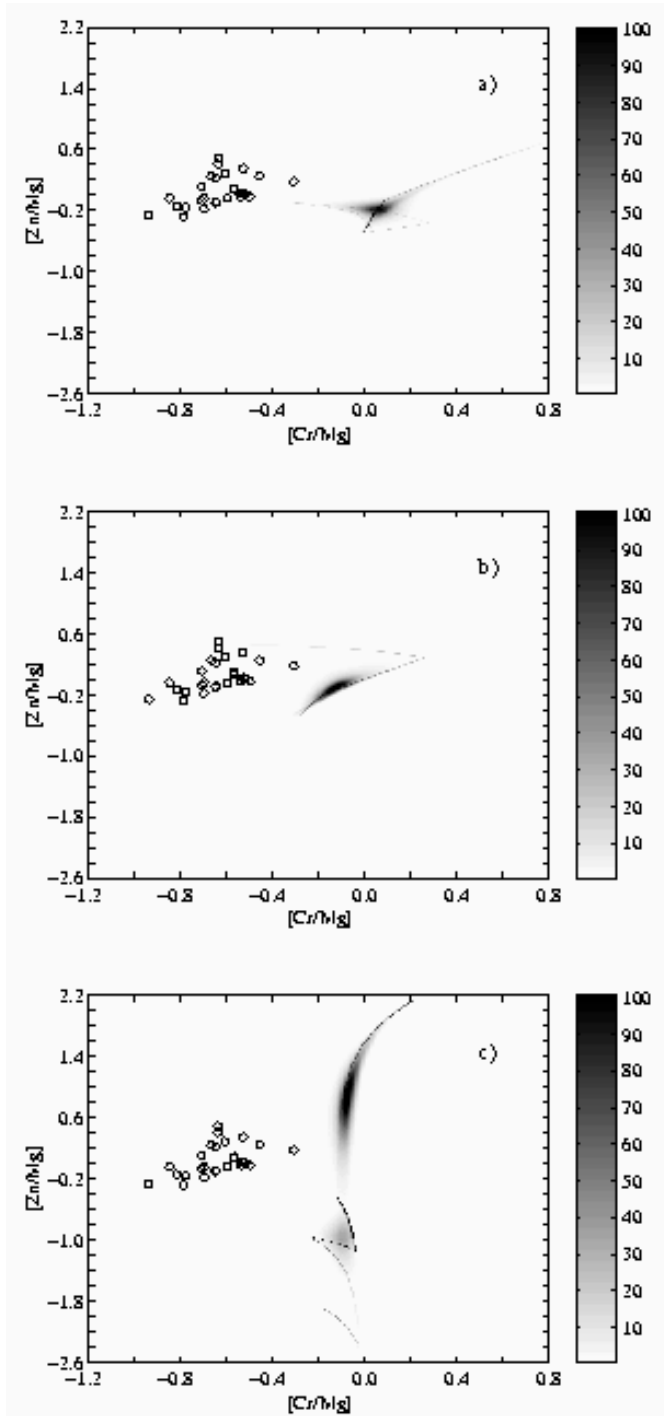}}
 \caption{Density functions of low-mass stars in the [Zn$/$Mg]--[Cr/Mg] plane. Symbols, see Fig. \ref{TiCaMg_df}. {\bf a)} Yields by WW95. {\bf b)} Yields by UN02. {\bf c)} Yields by CL04. The density function is multiplied by a factor of $3$ in order to enhance the contrast.}
 \label{ZnCrMg_df}
\end{figure}
 
\par 

We also note in passing that the recently found extremely metal-poor 
dwarf/subgiant star HE 1327-2326 ([Fe$/$H]$\simeq -5.4$, 
Frebel et al. 2005\nocite{fetal05}) has Na$/$Mg$/$Al ratios quite 
different from those of the extremely metal-poor giant HE 0107-5240 
([Fe$/$H]$\simeq -5.3$, Christlieb et al. 2004\nocite{cetal04}). 
These different results are not compatible with any of the three sets 
of yields explored here, and seem to support the conclusion that 
additional scatter of unknown origin is contributing, either from 
supernovae or from other sites of nucleosynthesis.

\subsection{$A/A$ diagrams for the iron-group elements}
\noindent
In Fig. \ref{MnCrFe_sc} we display the $A/A$ diagrams for Mn$/$Cr$/$Fe with the three different
sets of yields. The synthesis of the light elements of the iron peak are usually
ascribed to explosive silicon burning. It is seen that the predicted Cr$/$Fe ratios seem 
systematically too high by about $0.2-0.6$ dex when the model calculations are 
compared with the observed abundances. This discrepancy for the most metal-poor 
stars was also noted by Argast et al. (2000). On the other hand, 
the calculated Mn$/$Fe ratios turn out approximately as observed 
when the yields of WW95 and of CL04 are used, while they are too low 
by $0.8$ dex with the yields of UN02. 

\par

The $A/A$ diagrams for Co$/$Cr$/$Fe (Fig. \ref{CoCrFe_df})
show significantly too small Co$/$Fe ratios with the yield
sets of WW95 and UN02, while they come
out about right for the CL04 yields. In the latter
case, however, the scatter now is too large. 

\par

The Ni$/$Cr$/$Mg and Zn$/$Cr$/$Mg diagrams (see Fig. \ref{ZnCrMg_df}) 
are very similar and display the right order of 
magnitude of the Ni$/$Mg and Zn$/$Mg ratios for all sets of yields, 
although the CL04 yields again predict ranges which are much too 
large. The Cr$/$Mg ratios are too high, similar to the Cr$/$Fe ratios. 

\par

We note in general that a lower Cr yield would improve the agreement 
with the A/A diagrams for the iron-group elements. At least as probable as a reason
for the discrepancy, however, is the presence of overionization of atomic Cr 
in the atmospheres of Extreme Pop II stars which would lead to systematically 
underestimated Cr abundances. It has earlier been proposed on observational 
grounds (see, e.g., Fran\c{c}ois et al. 2004) that the Fe yield 
of WW95 is too high for their sub-solar metallicity models. A lowering 
of the Fe yield would increase the discrepancy in our $A/A$ diagrams 
even more. It is moreover not clear whether there are physical reasons to 
adopt such a reduction for iron only, and not for the other elements of the 
iron group.

\subsection{A comparison with homogeneous models}
\noindent
Francois et al. (2004) have explored a homogeneous model of the chemical
evolution of the early Galaxy, as compared with the Cayrel et al. (2004) 
observations. They first adopted SN II yields from WW95, 
although those of SN models with initial solar composition and not, as we have 
preferred, those with zero metallicity. The model predictions were also 
renormalized, sometimes by considerable amounts, to fit the abundances in the Sun. The
predicted abundances give too small [Cr$/$Fe] by about $0.5$ dex as 
compared with the observations, which is
inconsistent with our results above. For [Mn$/$Fe] Francois et al. (2004)
also found somewhat low values while their predicted 
[Co$/$Fe] and [Zn$/$Fe] are significantly too high (by typically $0.5$
dex) for the most metal-poor stars. This is in considerable disagreement with
our results, and may be explained as a consequence of the choice of initial 
solar abundances for the SN models as well as of the bias in their model
of the most massive SNe contributing to the most metal-poor stars, as will
be discussed immediately below.

\begin{figure}
 \resizebox{\hsize}{!}{\includegraphics{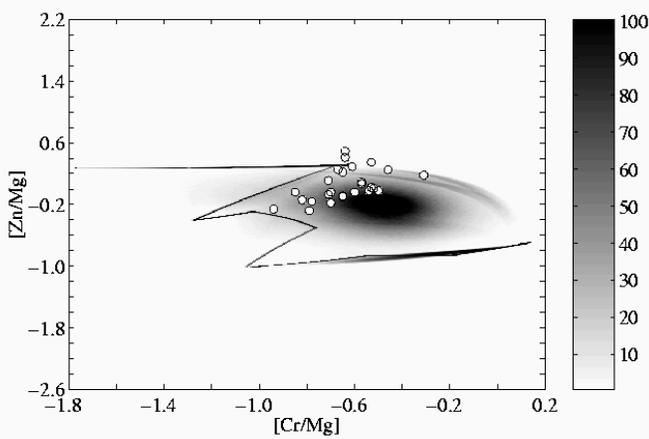}}
 \caption{Density function of low-mass stars in the [Zn$/$Mg]--[Cr/Mg] plane, using the empirically determined yields by Fran\c{c}ois et al. (2004). The density function is multiplied by a factor of $20$ in order to enhance the contrast. Symbols, see Fig. \ref{TiCaMg_df}.}
 \label{ZnCrMgF04_df}
\end{figure}

\par

The trends seen at very low [Fe$/$H] in [$A/$Fe] vs. [Fe$/$H] in 
the Fran\c{c}ois et al. (2004) results are interesting. In their model 
[Fe$/$H] is a direct function of time and and the relatively steep
gradients in [$A/$Fe] found for certain elements $A$ reflect the gradual
"switching on" of SNe in order of decreasing initial SN mass, within the 
first tens of million years in the model evolution. The fact that the 
yields of element $A$ change
when the SN mass decreases from $100$ to $10$ solar masses is thus mapped
directly onto the model track in the [$A/$Fe]-[Fe$/$H] diagram. This
effect cannot be obtained with our scenario, where the first 
stars exploding as SNe are picked randomly from the IMF, irrespectively of
their lifetimes. In fact, the assumption that the most massive SNe 
explode first in the Galactic gas, and that they contribute their fresh elements
to the gas before stars in the next mass range explode, requires
mixing times in the star-forming gas of characteristically 
$10$ Myr or less, which seems unrealistic. 

\par

In order to improve the fit between their model tracks in the $A/$H diagrams
and observations, Francois et al. (2004) have adjusted their yields as
a function of SNe mass. These adjustments are considerable, in particular
for elements like Mg and Cr, and drastically varying with SN mass. They
would lead to strong effects for the early chemical evolution of inhomogeneous
models, as seen in Fig. \ref{ZnCrMgF04_df}, which displays the effects in the 
Zn$/$Cr$/$Mg diagram with our model, however, with the Francois et al. adjusted 
yields (cf. Fig. \ref{ZnCrMg_df}). We note the extended spurs, mapping the very 
strong mass dependence of the yield ratios. This will lead to a significantly 
increased scatter in $A/A$ diagrams produced by inhomogeneous models, a scatter 
that is much greater than that observed for the first stars.

\subsection{Clustered star formation}
\noindent
It is often argued that the generally small star-to-star scatter recently observed for 
many abundance ratios is a result of an averaging of the SN ejecta in a well-mixed 
ISM. As discussed in Sect. \ref{ahdiagrams}, a small scatter can be obtained
in an inhomogeneous and poorly-mixed ISM as well (see, e.g., Fig. \ref{CaMgH_df}), 
depending on the details of the stellar yields. However, if star formation is strongly 
clustered, the ejecta of many SNe of different masses may mix before new stars are 
formed out of the enriched gas. The trends observed in several of the $A/$H diagrams 
would then be due to a metallicity-dependence of the IMF-averaged SN yields 
(``metallicity-enhancement effect''), rather than a mass-dependence of the 
individual SN yields (``stellar-mass effect''). A strong clustering would 
also wipe out the existence of SSSs in the otherwise much more robust $A/A$ diagrams. 

\par
 
In order to check whether clustered star formation will suppress the formation of stars enriched by a single or a few SNe to such a large extent, we randomly distributed a number of star-forming regions in a model box. Each region was given a certain size $V_{\mathrm{mix}}$, depending on the number $k_s$ of enriching SNe within the region. When a star-forming region overlapped with earlier formed regions, a number of model low-mass stars were allowed to form from the gas enriched by these earlier regions. This scenario is analogous to that of Oey (2000). The number of star-forming regions enriched by a certain number $k_s$ of SNe appears to follow a universal power-law proportional to $k_s^{-2}$, all the way down to $k_s=1$ (see Oey et al. 2004\nocite{okp04} and references therein). Assuming that this relation holds also in the young universe, we find that the number $n_k$ of stars enriched by $k$ SNe increases with decreasing $k$ (at least for $k\lesssim 35$), similarly to the result found in Paper~I for unclustered star formation (Paper~I, Fig. 5). Hence, there should presumably exist a fair number of stars only enriched by a single or a few SNe, also in an ISM with clustered star formation. In the light of this result we argue that scatter and structures, including SSSs, may be expected in the $A/A$ diagrams when observed with sufficient accuracy, independently of whether star formation is clustered or not. In Paper~III\nocite{paperiii}, the problems of scatter related to clustering (the averaging of many SNe), cooling, and mixing of the ISM are discussed in more detail.

\section{Conclusions}\label{conclusions}
\noindent
We have explored the predicted abundance ratios for very metal-poor stars from a 
stochastic model of early Galactic evolution (Paper I\nocite{paperi}) for different 
sets of SN yields (WW95, UN02, and CL04), and compared the predictions with observations. 
As regards the $A/\mathrm{H}$ diagrams, we find reasonable agreement between 
predictions and observations for Si and Ca, although the sloping trends are in general 
more pronounced in the simulations (WW95 and UN02), leading to Ca/Mg ratios which are 
too high for the most metal-poor stars. The predicted distribution of stars also agree 
fairly well with observations in the [Ti/Mg]--[Mg/H] plane, however, only 
when the yields by WW95 are used. The presence in the simulation results, as well as in 
those by Argast et al. (2000\nocite{argast00}), of a population of stars with very low 
Mg/Fe ratios are found neither in the survey of Cayrel et al. (2004\nocite{cayrel04}), 
nor of Cohen et al. (2004\nocite{cohen04}). The small scatter in abundance ratios, also 
among the most metal-poor stars, may be the result of cosmic selection effects in 
contributing SN masses, although other phenomena may also be significant, such as 
different degrees of mixing of the ISM with infalling unprocessed gas, or the effects 
of accretion of well-mixed interstellar gas onto extremely metal-poor stars. 

\par

The distribution of stars in the more robust $A/A$ diagrams, where various abundance 
ratios of heavy elements are plotted relative to each other, show rather good 
agreement for the $\alpha$-elements Mg, Si, and Ca, as regards mean values, 
trends and scatter (taking observational errors into consideration). However, 
the yields by UN02 and CL04 produce too low Ti/Mg ratios, as is also suggested by the 
discrepancy between predictions and observations in the $A/\mathrm{H}$ diagrams. 
For the odd-even elements Na and Al, the SN yields tend to give too small abundances 
in general, and the predicted scatter is lower than the observed one. For the iron-group 
elements the calculated yields when used in our models are only moderately successful 
in predicting the relative abundance ratios and their scatter. The sensitivity of the 
appearance of the $A/A$ diagrams to the particular yields illustrates, however, the 
possibility of these observational data to verify, or disprove, the validity of the 
yields adopted.

\par

The verification of the existence of ``spurs'' or ``Single Supernova Sequences'' in 
the $A/A$ diagrams, delineating stars with matter essentially affected by only one 
SN, is a considerable observational challenge for the future. Sample sizes of 
hundreds of very metal poor stars will have to be studied at very high spectroscopic 
accuracy  and homogeneity in order to find these. However, if found they will 
give very valuable information about the first generation of stars and SNe. The 
spurs, and other structures in the diagrams, may be expected to be present, even if 
clustered star formation occurs instead of uncorrelated formation of single stars, 
as assumed in our model. If not found in future accurate surveys, it may indicate 
the existence of early element production by supernovae with effective  mixing of 
the gas before any formation of low-mass stars occurred.

\begin{acknowledgements}
\noindent
We would like to thank Mike Edmunds for valuable discussions and Roger Cayrel 
for giving us a preprint of Cayrel et al. (2004) in advance. We also thank Anna Frebel 
and Norbert Christlieb for letting us quote their abundance results for HE 1327-2326. 
Andreas Korn and the anonymous referee are thanked for valuable comments on the 
manuscript. The work has been supported by the Swedish Research Council.
\end{acknowledgements}

\bibliographystyle{bibtex/apj}
\bibliography{ref}

\begin{appendix}
\section{Derivation of $f_{AB,CD}(x,y)$}\label{app_derivation}
\noindent
The stochastic theory presented in KG (Karlsson \& Gustafsson 2001\nocite{kg01}) and in Paper~I (Karlsson 2005a\nocite{paperi}) is essentially built on the idea that transformations between the distribution of supernova masses and the distribution of elements in subsequently formed stars can be found, via the yields. Here, we shall present a short summary of the results given in KG. Let us first discuss the transformation between random variables in the one-dimensional case.

\par
  
Let the IMF be a continuous function of stellar mass $m$ such that $\phi=\phi(m)=\phi_0m^{-\alpha}$, normalized as

\begin{equation}
\int\limits_{m_{l}}^{m_{u}}\phi(m) \mathrm{d}m = 1.
\label{imf_norm}
\end{equation}

\noindent
Regarding the stellar mass as a random variable $\mathcal{M}$, the distribution of masses over the range $[m_{l},m_{u}]$ is determined by the IMF and described by the probability density function $f_m(m)=\phi(m)$. Similarly, the yield of an element $x=p(m)$, which we assume is a continuous function of stellar mass, will be distributed according to the random variable $X=p(\mathcal{M})$. The probability that stars generate a yield which is smaller than or equal to a certain value, i.e. $X\le x=p(m)$, for monotonically increasing functions $p(m)$, is given by the distribution function 

\begin{eqnarray}
F_X(x) & = & P(X\le x) = P(p(\mathcal{M})\le x) \nonumber\\ 
& = & P(\mathcal{M}\le p^{-1}(x)) = F_m(p^{-1}(x)).
\label{Fox}
\end{eqnarray}

\begin{figure}
 \resizebox{\hsize}{!}{\includegraphics{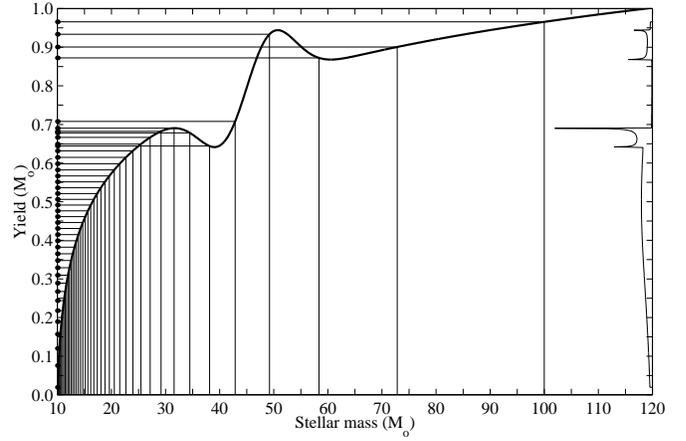}}
 \caption{An illustration of the transformation between random variables. The figure shows a hypothetical SN yield $p$ as a function of stellar mass (thick line). The thin vertical lines are distributed according the Salpeter IMF. The corresponding horizontal lines and the dots on the left hand side of the figure are distributed in yield-space according to the random variable $X=p(\mathcal{M})$. The probability density function of $X$, given by Eq. (\ref{fofp}), is shown on the right hand side of the figure.}
 \label{illustr}
\end{figure}

\noindent
The general expression for the density function (see Fig. \ref{illustr}) is then governed by 

\begin{eqnarray}
f_X(x) & = & \left| \frac{\mathrm{d}}{\mathrm{d}x}F_X(x) \right| = \left| \frac{\mathrm{d}}{\mathrm{d}x}F_m(p^{-1}(x)) \right| \nonumber\\
& = & \sum\limits_{i=1}^{j} f_{m}(p_i^{-1}(x)) \left| \frac{\mathrm{d}p_i^{-1}}{\mathrm{d}x} \right| = \sum\limits_{i=1}^{j} \frac{\phi(p_i^{-1}(x))}{\left| p_i'(p_i^{-1}(x)) \right|},
\label{fofp}
\end{eqnarray}

\noindent
where the yield is divided into $j$ parts such that $p=p_1+...+p_j$ and each function $p_i$ is monotonic and equivalent to $p$ on the open subinterval $]m_{i-1},m_{i}[$ and zero elsewhere. The $m_i$'s, $i=1,...,j-1$ are real roots to $p'(m)$ and $m_0$, $m_j$ are the end-points. Similarly to the $p_i$'s the functions $p_i'\equiv dp_i/dm$ on $]m_{i-1},m_{i}[$ and zero elsewhere. 

\par 

In the multi-dimensional case the transformation looks somewhat different but the idea is the same. The total ratio of two elements $A$ and $B$ produced and ejected from $k$ SNe can be regarded as a yield ratio in $k$ dimensions and is written as (cf. Eq. (26) in Paper~I\nocite{paperi})

\begin{equation}
y^{\xi}_{k(AB)}(m_1,...,m_k)=\frac{\sum\limits_{i=1}^k \xi_i 
p_{A}(m_i)}{\sum\limits_{i=1}^k \xi_i p_{B}(m_i)},
\label{p_ratio}
\end{equation}
 
\noindent
where $p_{A}(m_i)$ and $p_{B}(m_i)$ are the normal one-dimensional yields for a SN of progenitor mass $m_i$. The weights $\xi_i$ on each yield is a measure of the dilution of the ejecta due to mixing with the ambient ISM and was shown in Paper~I to be $\xi_i\equiv 1/M_i$, where $M_i$ is the mixing mass of the $i$th SN. The mixing masses are distributed according to the probability density function $f_{M_k}$ (see Appendix \ref{app_nkfmk}). The weights are thus distributed according to the random variable $\Xi_k$, described by the density function $f_{\Xi_k}$. Using the result in Eq. (\ref{fofp}), $f_{\Xi_k}$ is can be written as
\begin{equation}
f_{\Xi_k}=\frac{1}{\xi^2}f_{M_k(1/\xi)}.
\label{f_xi}
\end{equation}

\noindent
Note that the distribution of mixing masses depends on $k$, the number of enriching SNe. Therefore, the weights depend on $k$ as well. As shown in KG the distributions of stars in the $A/A$ diagrams are insensitive to the distribution of the weights since probability density functions of ratios between two arbitrary, equally distributed random variables have similar shapes and are strongly peaked towards unit ratio. The presence of large-scale mixing only wipes out the structures on the smallest scales in these diagrams.

\par

Using the definition of $y^{\xi}_{k(AB)}$ in Eq. (\ref{p_ratio}), the random variable describing the total ratio between two elements is then given by

\begin{eqnarray}
X^{\xi}_{k(AB)} & = & [y^{\xi}_{k(AB)}(\mathcal{M}_1,...,\mathcal{M}_k)] \nonumber\\
& = & \log\left(\frac{\sum\limits_{i=1}^{k}\Xi_{k,i} p_{A}(\mathcal{M}_i)}{\sum\limits_{i=1}^{k}\Xi_{k,i}p_{B}(\mathcal{M}_i)}\right)-\log(A/B)_{\odot},
\end{eqnarray}

\noindent
where the superscript $\xi$ is just a reminder that these functions also depend on the amount of dilution (described by $\Xi_k$. Recall that $\mathcal{M}$ is the random variable of stellar masses. The common astrophysical notation introduced above, where the abundance ratios are measured on a logarithmic scale relative to the Sun could as well have been introduced later as a transformation operating on the linear density functions (see Eqs. (A.16) and (A.17) in KG). 

\par

The joint distribution of the two random variables $X^{\xi}_{2k(AB)}$ and $Y^{\xi}_{2k(CD)}$ for the elements $A$, $B$, $C$, and $D$ describes how stars, enriched by $k$ SNe, would populate the [$A/B$]--[$C/D$] plane. The distribution is given by the two-dimensional function $f_{k(AB,CD)}(x,y)$, where $x=[A/B]$ and $y=[C/D]$. In its present form, Eq. (\ref{fofp}) cannot be used to calculate multi-dimensional probability functions. Instead, we may use the alternative expression (see e.g. Papoulis 1991\nocite{papoulis91})

\begin{eqnarray}
f_{k(AB,CD)}(x,y) & = & \int...\int_{\Delta D^k_{xy}} \phi(m_1) 
\times... \nonumber\\
& \times & \phi(m_k)\mathrm{d}m_1...\mathrm{d}m_k.
\label{f_kXY}
\end{eqnarray}   

\noindent
The integrand is formed by a simple product of IMFs and does not contain any information about where the stars are distributed in yield-space. One of the main results in KG is that the chemical abundances patterns in the $A/A$ diagrams are determined by the differential integration regions $\Delta D^k_{xy}$. For every point $(x,y)$ in the [$A/B$]--[$C/D$] plane, $\Delta D^k_{xy}$ determines whether the density of stars is different from zero or not. The integrand only partly contributes to the total density. The integration region is formed by the intersection in $m$-space between $\Delta D^k_{x}$ and $\Delta D^k_{y}$, i.e., $\Delta D^k_{xy}=D^k_{x} \bigcap \Delta D^k_{y}$. These regions are in turn defined as 

\begin{equation}
\left\{ \begin{array}{ll}
\Delta D^k_{x}=\{(m_1,...,m_k);~x<[y^{\xi}_{k(AB)}]\le x+\mathrm{d}x\}\\
\Delta D^k_{y}=\{(m_1,...,m_k);~y<[y^{\xi}_{k(CD)}]\le y+\mathrm{d}y\},
\end{array} \right.
\end{equation}

\noindent
where the yield ratios are given by Eq. (\ref{p_ratio}), although on a logarithmic scale. Thus, for any differential region in $xy$-space with a non-zero density there is at least one region in $m$-space for which the combination of masses (and dilution factors) simultaneously produces abundance ratios $x=[A/B]$ and $y=[C/D]$.  

\par

The total probability density function of ($X_{AB},Y_{CD}$), describing the distribution of low-mass stars, enriched by any number of SNe from one to $n$, is then formed by a sum of the individual density functions $f_{k(AB,CD)}(x,y)$, given by Eq. (\ref{f_kXY}), such as

\begin{equation}
f_{AB,CD}(x,y)=\sum\limits_{k=1}^{n} w_k \times f_{k(AB,CD)}(x,y).
\label{finalexpr}
\end{equation}

\noindent
The normalization weight $w_k$ is the fraction of stars enriched by $k$ SNe such that $w_k=n_k/\sum_{k=1}^n n_k$, where $n_k$ is the number (kpc$^{-3}$) of stars enriched by $k$ SNe. The expression for $n_k$ can be found in Appendix \ref{app_nkfmk}. 

\par

In Eq. (\ref{finalexpr}), the summation is terminated at $n$, a high number of enriching SNe. However, for observational comparison, it would be more correct to let the probability density functions be terminated at a certain metallicity. Formally, this is achieved by collapsing the corresponding three-dimensional density function $f_{AB,CD,Z}(x,y,z)$ in $z$, where $z$ is the metallicity measured by, e.g., [Fe$/$H]. Analogously to the two-dimensional probability density functions described above, $f_{AB,CD,Z}$ is a sum of partial functions $f_{k(AB,CD,Z)}$ which in turn are formed by integration over the regions $\Delta D^k_{xyz}=\Delta D^k_{x} \bigcap \Delta D^k_{y} \bigcap \Delta D^k_{z}$, and where

\begin{equation}
\Delta D^k_{z}=\{(m_1,...,m_k);~z<[y_{k(Z)}]\le z+\mathrm{d}z\}.
\end{equation} 

\noindent
The expression for $y_{k(Z)}(m_1,...,m_k)$ is given by Eq. (26) in Paper~I and describes the possible metal content (measured relative to hydrogen) of a star enriched by $k$ SNe. It is completely analogous to the expression for the abundance ratio of two heavier elements in Eq. (\ref{p_ratio}). The marginal density function $f_{AB,CD}(x,y)$ is then given by 

\begin{equation}
f_{AB,CD}(x,y)=c_Z\int\limits_{Z_l}^{Z_u} f_{AB,CD,Z}(x,y,z) \mathrm{d}z,
\label{fterminated}
\end{equation}

\noindent
where $Z_l\rightarrow -\infty$ is the lowest possible metallicity and $Z_u$ is the termination metallicity (e.g. $z_u=[\mathrm{Fe}/\mathrm{H}]=-2.5$). the factor $c_Z$ is a normalization constant.

\section{Renormalization of  $f_{AB,CD}(x,y)$}\label{app_renorm}
\noindent
We shall briefly discuss a way of renormalizing the probability density functions for comparison with biased observational samples. Observations of stellar populations intended to map the Galactic chemical evolution are often designed such that the stars are evenly spread over the metallicity interval in log-space. However, density functions given by Eqs. (\ref{finalexpr}) and (\ref{fterminated}) describe the distribution of a volume-limited sample and show an approximately exponential increase in number of stars per metallicity bin for increasing metallicity (see Fig. 6 in Paper~I). Renormalization to constant density per metallicity bin is done by again integrating the three-dimensional density function $f_{AB,CD,Z}(x,y,z)$, this time with a weighting function $w(z)$ such that 

\begin{eqnarray}
f^{\mathrm{renorm}}_{AB,CD}(x,y) & = & \int\limits^{\infty}_{-\infty} w(z) f_{AB,CD,Z}(x,y,z) \mathrm{d}z\nonumber\\
& = &  \int\limits^{Z_u}_{Z_l} w(z) f_{AB,CD,Z}(x,y,z) \mathrm{d}z. 
\label{frenorm1}
\end{eqnarray}

\noindent
The weighting function $w(z)$ is essentially the inverse to the function $f_Z$, describing the distribution of stars as a function of metallicity (see Eq. (28) in Paper~I) This function can also be derived from $f_{AB,CD,Z}(x,y,z)$ by collapsing the two dimensions in $x$ and $y$ such that

\begin{equation}
w(z)=\left( L_Z\int\limits^{\infty}_{-\infty}\int\limits^{\infty}_{-\infty} f_{AB,CD,Z}(x,y,z) \mathrm{d}x \mathrm{d}y \right)^{-1},
\label{weight}
\end{equation} 

\noindent
i.e., $w(z)$ is, except for the normalization factor $L_Z^{-1}=1/(Z_u-Z_l)$, the inverse of the marginal density function $f_Z$ of $Z$. Using Eq. (\ref{weight}), $f^{\mathrm{renorm}}_{AB,CD}$ can then be rewritten as

\begin{eqnarray}
f^{\mathrm{renorm}}_{AB,CD}(x,y) & = & \frac{1}{L_Z} \int\limits^{Z_u}_{Z_l} f_{AB,CD,Z}(x,y,z)/f_Z(z) \mathrm{d}z \nonumber\\
& = & \frac{1}{Z_u-Z_l}\int\limits^{Z_u}_{Z_l} f_{AB,CD|Z}(x,y|z) \mathrm{d}z. 
\label{frenorm2}
\end{eqnarray}

\noindent 
The function $f_{AB,CD|Z}(x,y|z)$ is the conditional probability density function of $X_{AB}$ and $Y_{CD}$ given that $Z=z$. It describes the distribution of stars in the [$A/B$]--[$C/D$] plane for a given metallicity. Integrating over metallicity will then give the renormalized density function that we seek.

\section{Expressions for $n_k$ and $f_{M_k}$}\label{app_nkfmk}
\noindent
Assuming that the probability of finding a region enriched by $k$ SNe at time $t$ is given by Eq. (\ref{poisson}), the total number density of stars formed in such regions in the time interval $[t,t+\mathrm{d}t]$ is $w_{\mathrm{ISM}}(k,t)\psi(t)\mathrm{d}t$, where $\psi(t)$ is the SFR per unit time and unit volume. The number of still-existing stars (enriched by $k$ SNe) per unit volume formed up to $\tau_{\mathrm{G}}$, the age of the Galaxy, is then given by the integral

\begin{equation}
n_{k} = \int\limits_{0}^{\tau_\mathrm{G}} a_{\star}(t)w_{\mathrm{ISM}}(k,t) \psi(t)\mathrm{d}t,
\label{nk}
\end{equation}

\noindent
where $a_{\star}(t)$ denotes the fraction of a stellar generation formed at $t$ that still exists today and is expressed as

\begin{equation}
a_{\star}(t)=\int\limits_{m_{l}}^{g_{\tau}^{-1}(\tau_{\mathrm{G}}-t)}\!\!\!\!\!\!\!\!\!\phi(m)\mathrm{d}m, 
\label{aexist}
\end{equation} 

\noindent
where $\phi(m)$ denotes the normalized IMF, $m_l$ is the mass of the least massive stars, and $g_{\tau}^{-1}(\tau_{\mathrm{G}}-t)$ is the mass of the most massive, still-surviving stars formed at time $t$. In the early Galaxy, when most of the stars with [Fe$/$H] $<-2.5$ were formed, $a_{\star}$ changes very little. In the extremely metal-poor regime we may therefore approximate the integral in Eq.~(\ref{aexist}) with a constant such that $a_{\star}=a_{\mathrm{LMS}} $, where $a_{\mathrm{LMS}}$ is the fraction of low-mass stars with lifetimes greater than the age of the Galaxy. In Paper~I\nocite{paperi}, this fraction was estimated to $a_{\mathrm{LMS}}=0.835$.

\par

Assuming homogeneous mixing, the amount of dilution of ejected SN material is proportional to the mass of the interstellar material inside the mixing volume associated with the SN. In the general case, for a time-dependent density $\rho(t)$, the total mass $M$ inside the mixing volume does not only depend on the size of $V_{\mathrm{mix}}$ but also on the time when $V_{\mathrm{mix}}$ was created by the SN explosion. Let $\tau_{V}$ be the time that has elapsed since the creation of $V_{\mathrm{mix}}$. Thus, $\tau_{V}$ denotes the age of $V_{\mathrm{mix}}$ and is a measure of its size. If the SN exploded at time $t-\tau_{V}$, the mass $M$ is given by the two-dimensional function $g_{M}$, where

\begin{equation}
g_{M}(t,\tau_{V}) = 
\left\{ \begin{array}{ll}
\!\!\int\limits_{t-\tau_{V}}^{t}\!\!\!\dot{V}_{\mathrm{mix}}(t'-t+\tau_{V})\rho(t')\mathrm{d}t', & t\ge \tau_{V} \\
0, & t<\tau_{V}.  \end{array}
\right.
\label{vrmass}
\end{equation} 

\noindent
Note that the mass of enriched gas only grows by mixing of the {\it surface} of $V_{\mathrm{mix}}$ with the surrounding material. The mixing mass cannot change, and in particular not decrease, due to density changes {\it within} $V_{\mathrm{mix}}$. Hence, the term $V_{\mathrm{mix}}\dot{\rho}$ is excluded in the integral in Eq. (\ref{vrmass}).

\par

\begin{figure}[t]
 \resizebox{\hsize}{!}{\includegraphics{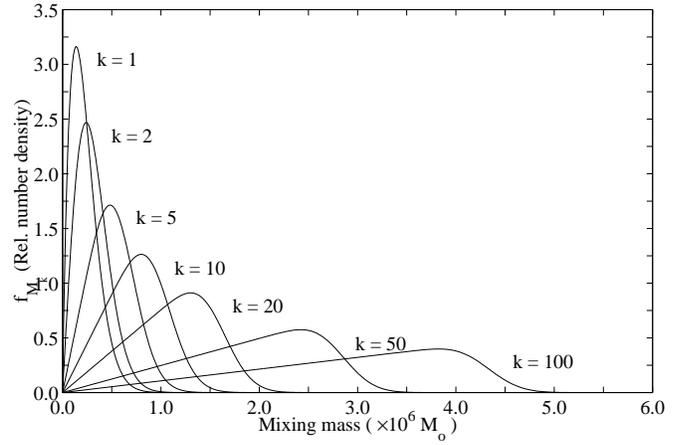}}
 \caption{The distribution of mixing masses in regions enriched by $k=1,2,5,10,20,50,$ and $100$ SNe as given by Eq.~(\ref{derivedfm}) for a constant SFR (see Model A in Paper~I).}
 \label{Mmix_fig}
\end{figure}

Now, at time $t\ge \tau_{V}$, mixing volumes of size $V_{\mathrm{mix}}(\tau_{V})$ appear at a rate $u_{\mathrm{SN}}(t-\tau_{V})$, which is a measure of how many mixing volumes there are of this size at time $t$. Thus, the probability that such a mixing volume overlaps a region already enriched by $k-1$ SNe at time $t$ is proportional to $V_{\mathrm{mix}}(\tau_{V}) w_{\mathrm{ISM}}(k-1,t) u_{\mathrm{SN}}(t-\tau_{V})$. The probability of forming a still-existing star in this region in the time interval $[t,t+\mathrm{d}t]$ and age interval $[\tau_{V},\tau_{V}+\mathrm{d}\tau_{V}]$ is then proportional to $a_{\mathrm{SN}}a_{\star}(t)V_{\mathrm{mix}}(\tau_{V}) w_{\mathrm{ISM}}(k-1,t)\psi(t-\tau_{V}) \psi(t)\mathrm{d}t\mathrm{d}\tau_{V}$, where we have assumed that $u_{\mathrm{SN}}=a_{\mathrm{SN}}\psi$. Integrating this expression over the region $\Delta D_{M}$ in the $(t,\tau_{V})$-plane gives the total probability $f_{M_k}$ of having a mixing mass of size $M$ in regions enriched by $k$ SNe. Thus, 

\begin{eqnarray}
f_{M_k} & = & c_k\!\!\!\!\int\limits_{\Delta D_{M}}\!\!\!\!a_{\star}(t)V_{\mathrm{mix}}(\tau_{V}) w_{\mathrm{ISM}}(k-1,t)\times \nonumber\\
 &\times & \psi(t-\tau_{V})\psi(t)\mathrm{d}t\mathrm{d}\tau_{V},
\label{generalfm}
\end{eqnarray}

\noindent
where the SN fraction $a_{\mathrm{SN}}$ has been included in the normalization constant $c_k$. The differential integration region $\Delta D_{M}$ is defined by the function $g_{M}(t,\tau_{V})$ in Eq.~(\ref{vrmass}) such that 

\begin{equation}
\Delta D_{M}=\{(t,\tau_{V});~M<g_{M}(t,\tau_{V})\le M+\mathrm{d}M\}.
\end{equation}   

\par 

Eq.~(\ref{generalfm}) may be simplified by assuming that the density of the ISM is kept constant by a continuous infall of pristine gas. The mixing mass $M$ is then directly proportional to the mixing volume such that 

\begin{equation}
M=\overline{\rho} V_{\mathrm{mix}}(\tau_{V})\equiv g_{M}(\tau_{V}),
\label{crmass}
\end{equation}

\noindent
where $\overline{\rho}$ is a typical density. Since the mass is determined by the size of $V_{\mathrm{mix}}$ alone, the integration over $\tau_{V}$ is redundant and Eq.~(\ref{generalfm}) reduces to a one-dimensional integral

\begin{equation}
f_{M_k} = \,\,\,\tilde{c}_kM\!\!\!\!\!\!\!\int\limits_{g_M^{-1}(M)}^{\tau_{\mathrm{G}}}\!\!\!\!\!\! w_{\mathrm{ISM}}(k-1,t)\psi(t-\tau_{V})\psi(t)\mathrm{d}t,
\label{derivedfm}
\end{equation}

\noindent
where we have used Eq.~(\ref{crmass}) to write $V_{\mathrm{mix}}(\tau_{V})$ in terms of $M$, which can be taken outside the integral. The factors $a_{\mathrm{LMS}}$ and $1/\overline{\rho}$ have been included in the normalization constant $\tilde{c}_k$. Recall that $\tau_{\mathrm{G}}$ denotes the age of the Galaxy. Since mixing masses of size $M=g_{M}(\tau_{V})$ do not exist at times $t<\tau_{V}$, a lower integration limit is introduced at $t=\tau_{V}$, making it, as indicated, a function of $M$. Mixing mass distributions for different number $k$ of enriching SNe are plotted in Fig. \ref{Mmix_fig}. The derivations of $n_k$ and $f_{M_k}$ can also be found in Paper~I\nocite{paperi}.

\end{appendix}

\end{document}